\documentclass[useAMS,usenatbib]{mn2e}
\usepackage{graphicx}
\usepackage{amssymb}
\usepackage{color}

\title[Stellar mass maps]{Resolved stellar mass maps of galaxies. I:
  method and implications for global mass estimates}

\author[Zibetti, Charlot \& Rix]{Stefano Zibetti$^{1}$\thanks{E-mail:
    zibetti@mpia.de}, St\'ephane Charlot$^{2}$, Hans-Walter Rix$^{1}$ \\
  $^{1}$Max-Planck-Institut f\"ur Astronomie, K\"onigstuhl 17, D-69117 Heidelberg, Germany\\
  $^{2}$Institut d’Astrophysique de Paris, CNRS, Universit\'e Pierre \& Marie Curie, 98 bis Boulevard Arago, 75014 Paris, France\\
}
\begin{document}
\bibliographystyle{mn2e}

\date{Accepted 2009 August 10. Received 2009 August 10; in original form 2009 April 1}

\pagerange{\pageref{firstpage}--\pageref{lastpage}} \pubyear{2009}

\maketitle

  \label{firstpage}

\begin{abstract}
  We introduce a novel technique to construct spatially resolved maps
  of stellar mass surface density in galaxies based on optical and
  near IR imaging. We use optical/NIR colour(s) to infer effective
  stellar mass-to-light ratios ($M/L$) at each pixel, which are then
  multiplied by the surface brightness to obtain the local stellar
  surface mass density. We build look-up tables to express $M/L$ as a
  function of colour(s) by marginalizing over a Monte Carlo library of
  50,000 stellar population synthesis (SPS) models by Charlot \&
  Bruzual (2007), which include a revised prescription for the TP-AGB
  stellar evolutionary phase. Moreover, we incorporate a wide range of
  possible dust extinction parameters. In order to extract reliable
  flux and colour information at any position in the galaxy, we
  perform a median adaptive smoothing of the images that preserves the
  highest possible spatial resolution.

  As the most practical and robust, and hence fiducial method, we
  express the $M/L$ in the $H$ band as a function of $(g-i)$ and
  $(i-H)$. Stellar mass maps computed in this way have a typical
  accuracy of 30 per cent or less at any given pixel, determined from
  the scatter in the models. We compare maps obtained with our
  fiducial method with those derived using other combinations of
  band-passes: {\em i)} mass maps based on the $M/L$ in NIR bands
  require one optical and one optical-NIR colour to avoid significant
  biases as a function of the local physical properties of a galaxy;
  {\em ii)} maps based on $M/L$ in $i$ band as a function of $(g-i)$
  only are generally in excellent agreement with our best optical-NIR
  set, except for extremely star forming and dust extincted
  regions. We further compute stellar mass maps using a model library
  identical to the previous one except for being based on older SPS
  models, which assume shorter-lived TP-AGB stars. The $M/L$ in the
  NIR inferred using these old models may be up to 2.5 times larger
  than the new ones, but this varies strongly as a function of colours
  and is maximal for the bluest colours.

  Finally, we compare total stellar mass estimates obtained by
  integrating resolved mass maps with those obtained with unresolved
  photometry. In galaxies with evident dust lanes, unresolved
  estimates may miss up to 40 per cent of the total stellar mass
  because dusty regions are strongly under-represented in the luminous
  fluxes.
\end{abstract}


\section{Introduction}\label{intro_sec}
Stellar mass may be the most fundamental parameter describing present
day galaxies. A decade ago \cite{gavazzi+96} and
\cite{gavazzi_scodeggio96} pointed out that the structure and star
formation history of disk galaxies are tightly linked with stellar
mass. \cite{scodeggio+02} extended this observation to all
morphologies. The advent of the Sloan Digital Sky Survey \citep{SDSS}
and improved stellar population models to estimate stellar masses,
ages and metallicities have allowed to put previous claims on a more
secure and detailed ground, establishing the dependence of structure,
star formation and chemical enrichment histories on stellar mass
\citep[e.g.][]{kauffmann+03b,tremonti+04,gallazzi+05}. More recently,
\cite{vandenbosch+08} have shown that stellar mass also is the main
parameter determining the properties of ``satellite'' galaxies, almost
irrespective of their parent dark matter halo mass or of the
halo-centric distance: in other words, stellar mass is by far more
crucial for predicting, or setting, galaxy properties than
environment.

\cite{bell_dejong00} pointed out that the mean stellar mass density of
a galaxy may be an even more basic parameter than the total stellar
mass in determining the stellar populations in spiral galaxies. This
conclusion has been confirmed by \cite{kauffmann+03b} and extended to
all morphological types using more than 100,000 galaxies from the
SDSS.  Stellar masses for most of these results were derived from
spectra or colours that were averaged across much of the stellar
bodies of the galaxies, arriving at a global mass-to-light ($M/L$)
ratio estimate.

In the light of these results, it is manifestly important to {\it i)}
test the accuracy of total stellar mass estimates and {\it ii)}
develop methods that actually map the stellar mass surface density
distribution in galaxies, rather than just inferring it from single
pass-band images rescaled by a uniform mass-to-light ratio.  Accurate
maps of stellar mass distribution of galaxies are also of fundamental
importance for dynamical studies aiming at disentangling the role of
secular vs. environmental induced evolution \citep[e.g.][Foyle et
al. 2009 in preparation]{kendall+08}.

Nowadays a wealth of multi-wavelength imaging is available for large
regions of the sky, covering from the UV to the optical, near and mid
IR. While most of the multi-wavelength work on galaxies has focused on
modeling the total or area-averaged spectral energy distribution, we
propose here to combine the photometric information on a
pixel-by-pixel level in order to retain the maximum spatial
resolution\footnote{Pioneering work in the pixel-by-pixel
      approach was already conducted by \cite{abraham+99} and
      \cite{conti+03}.}. With this method we primarily aim at
studying the stellar mass distribution within galaxies but we also aim
to study the dependence of the spectral energy distribution on stellar
mass density.

Past work has often attempted to map stellar mass distributions within
galaxies, resorting to NIR images as a proxy
\citep[e.g.][]{elmegreen_elmegreen84,rix_zaritsky95,seigar_james98,grosbol+04}.
More recently, \cite{kendall+08} have studied the spiral density waves
in M81 using three different methods to estimate stellar mass surface
density from $K_s$-band images alone, from 0.8 $\mu$m ($I$ band)
images with a pixel-by-pixel $M/L$ correction based on $B-V$ colours
\citep{bell_dejong01}, and based on Spitzer 3.6+4.5 $\mu$m imaging.

In this work we develop a rigorous method to derive spatially resolved
stellar mass density maps, based on sets of optical/NIR images of
galaxies. The basic idea is that at each position in a galaxy the
surface density of stellar mass is given by
\begin{equation}
  \Sigma_{M_*}(\alpha,\delta)=\Sigma_{\lambda}(\alpha,\delta) \Upsilon_{\lambda}(\alpha,\delta)
\end{equation}\label{ML_eq}
where $\Sigma_{\lambda}(\alpha,\delta)$ and
$\Upsilon_{\lambda}(\alpha,\delta)$ are the surface brightness and the
{\it effective} stellar mass-to-light ($M/L$) ratio in a pass-band of
effective wavelength $\lambda$. In turn, $\Upsilon_{\lambda}$ can be
expressed as a function of one or more colours, as measured at the
given location in the galaxy. By ``effective'' $M/L$ ratio we mean the
ratio between stellar mass and the light that reaches the observer, as
opposed to the light that is emitted and can possibly be absorbed by
the dust inside the galaxy.\footnote{Emission lines from the gaseous
  interstellar medium can affect the ``effective'' $M/L$ as well.  We
  will not take their contamination into account explicitly, but only
  implicitly by not using pass-bands that include the H$\alpha$
  emission.} In Section \ref{SPS_sec} we derive the recipes that allow
us to express $M/L$ as a function of colours, based on a Monte Carlo
library of last generation stellar population synthesis models
\citep[i.e. the 2007 version of the models by][]{BC03}, which also
include physically motivated prescriptions that account for dust
absorption. As we will show, in order to derive $M/L$ that are
accurate within $\approx 30$ per cent, colour(s) must be accurate at
0.1 magnitude or better. This requires for each resolution element
(pixel) a signal-to-noise ratio ($S/N$ hereafter) of $\gtrsim
20$. While typical imaging surveys, like the SDSS and medium-depth
observation in the NIR, can easily warrant such $S/N$ for the central
and brightest pixels of a galaxy, for lower surface brightnesses
($\mu_r\gtrsim 23$mag arcsec$^{-2}$, typically for $R\gtrsim
R_{\mathrm e}$) this condition cannot be met for individual
pixels. Image smoothing allows to enhance the $S/N$, but at expenses
of the effective image resolution. For this reason, we have developed
an adaptive median smoothing code ({\sc adaptsmooth}, Zibetti 2009 in
preparation) that preserves the maximum spatial information compatible
with minimum $S/N$ requirements. The image processing required to
compute stellar mass maps is detailed in Section \ref{imaging_sec}.

The goal of this paper is mainly to introduce this new method of
stellar mass density mapping based on optical and near-IR
imaging. Therefore, we limit our sample to only nine galaxies, which
span, however, a large range in morphologies. Our selection criteria
and the sources of imaging data are presented in Section
\ref{sample_sec}. Despite the small sample size, we find the very
interesting result that total stellar masses estimated from resolved
and from unresolved photometry differ quite significantly, as we
discuss in Section \ref{mass_bias_sec}. A summary with concluding
remarks and an outlook of future work is given in Section
\ref{sec_summary}.


\section{Methodology (I): stellar population synthesis models}\label{SPS_sec}
\subsection{Stellar $M/L$ from optical-NIR colours}
In this Section we combine stellar population synthesis (SPS) models
with simple prescriptions for dust attenuation to devise a set of
look-up tables that allow us to estimate the effective stellar
mass-to-light ratio given either one or two optical/NIR colours. As
mentioned above, by ``effective'' $M/L$ we mean the ratio between
stellar mass and the light that reaches the observer, after being
possibly absorbed by the dust within the galaxy.  To generate our
fiducial SPS models we adopt the 2007 version of Gissel \citep[CB07
hereafter]{BC03}, which includes revised prescriptions for the TP-AGB
evolutionary phase, following \cite{marigo_girardi07} and
\cite{marigo_girardi08}\footnote{Using simple stellar populations of
  different metallicities we have compared the $M/L$ ratios in red and
  NIR bands as a function of colours predicted by CB07 and
  \cite{maraston05} models. Although significant discrepancies up to a
  few tenths of a dex are seen, especially at the youngest ages, in
  the most relevant range of colours ($g-i>0.2$~mag, see
  Sect. \ref{sec:cc_space_models_obs}) no systematic offset is
  observed.}. An accurate modeling of this phase is of particular
relevance to correctly predict fluxes (hence colours and $M/L$) that
involve the wavelength range between 1 and 2.5 $\mu$m for stellar
populations of ages between 0.3 and 2 Gyr
\citep[e.g.][]{maraston05,bruzual07}. The emerging optical/NIR
colour(s) of the stellar population at a given position in a galaxy
are determined by a variety of factors, namely the star formation
history, the metallicity, the amount, spatial distribution and optical
properties of dust and the initial mass function (IMF) of stars. In
the following we will work under the assumption that the IMF is
universal and well described by \cite{chabrier03}. The systematic
effects on mass estimates induced by different choices of IMF
\citep[e.g.][]{bell_dejong01} are not of primary relevance here, as
long as the IMF can be considered uniform within a galaxy. This
assumption is suggested by Occam's razor, as we lack any way of
linking possible variation of IMF to local observables within a
galaxy.

All other relevant parameters, star formation history, metallicity,
dust, are expected to vary significantly from place to place in a
galaxy. To explore the effect of such differences on colours and
$M/L$, we build a Monte Carlo library of SPS models with the following
properties, as in \cite{kong+04} and \cite{dacunha+08}. Each stellar
population has a fixed metallicity, randomly chosen between 0.02 and 2
times solar, and a two-component star formation history (SFH),
consisting of a continuous, exponentially declining mode with random
bursts superimposed. We follow \cite{kauffmann+03} to parameterize
each SFH by a set of variables with a given prior probability
distribution. The effect of dust is modeled according to the formalism
introduced by \cite{charlot_fall00} to treat the differential
absorption by dust in the short-lived birth clouds and in the
interstellar medium (ISM). As in \cite{dacunha+08}, the two parameters
that govern dust absorption, the total effective $V$-band optical
depth $\hat{\tau}_V$ and the fraction contributed by dust in the
ambient ISM $\mu$, have $1-\tanh$ prior probability distributions,
such that $\hat{\tau}_V$ is approximately uniform over the interval
from 0 to 4 and drops exponentially to zero around $\hat{\tau}_V=6$,
while $\mu$ is approximately uniform over the interval from 0 to 0.6
and drops exponentially to zero around 1. For each model emerging
fluxes and stellar masses are combined to compute colours and
effective $M/L$ ratios.

To study how $M/L$ in a given band depends on colours we adopt a
marginalization approach. We bin models in the 1- or 2-dimensional
colour space we aim at studying, with a bin width
$\lesssim\sigma_{\rm{colour}}$, the typical observational
error. Specifically, we adopt a bin width of 0.05 mag in each colour
dimension. Within each bin we compute the median $M/L$ and the
logarithmic {\it r.m.s.} from all parameter combinations that lead to
those colours. This approach incorporates uncertainties from parameter
degeneracies or model simplifications. E.g. the assumption of a single
metallicity in a given SPS model may be unrealistic even on local
scales, but by marginalizing over a large number of random models with
different metallicities we indirectly take varying metallicity into
account.

In principle, more colours should provide better constrained stellar
populations and hence $M/L$; in practice, there are limitations: most
of the colour information is nearly degenerate, especially that from
adjacent bands like $r-i$ and $i-z$. We have found that using one more
colour just complicates the analysis by increasing its dimensionality
without reducing the $M/L$ uncertainty. More colours would only
provide improvements if the systematic flux uncertainties in the
models were smaller than 10 per cent, which current models do not
reach yet. For these reasons we limit our study to two colour indexes
at most.

For practical reasons, we only consider broad pass-bands with
extensive existing data sets: the SDSS ($u,~g,~r,~i$ and $z$) and the
$J,~H,~K$ filters in the NIR. Among them, we discard the $u$ band
because of the typically low S/N in SDSS and the very high dust
attenuation that can lead to almost complete obscuration over a
significant area of a galaxy. We further discard the $r$ band because
of the strong local contamination by $H\alpha$ emission in HII
regions: typical equivalent widths of a few hundreds \AA ngstroms in
those regions would lead to overestimate the stellar emission by up to
0.5 mag in $r$. Among the three NIR bands we focus on the $H$ as our
reference band, but we note that any conclusion we present regarding
this band applies almost identically to $J$ and $K$ as well.

It is well known from previous studies (e.g. \citealt{rix_zaritsky95}
appendix B, or \citealt{bell_dejong01}) that in NIR bands $M/L$
variations as a function of stellar population parameters are smaller
than at shorter wavelengths, because the bulk of long-lived low-mass
stars dominate the emission in the NIR. Moreover, dust extinction is
lower at longer wavelengths. Thus, for our fiducial mass
reconstruction method it is natural to choose a NIR band (namely the
$H$ band) as the ``luminance'' band, whose surface brightness we want
to convert to stellar mass surface density. As for the colour space
where we map $M/L$ we choose $(g-i)$-$(i-H)$, because these colours
provide the largest wavelength leverage and thus the highest
sensitivity to stellar population and dust properties. This large
leverage also minimizes the effect of any systematic uncertainties in
the photometric calibration of SPS models or of the imaging data: 10
per cent flux uncertainties are negligible for a colour range of 2
mag, but not for a range of a few 0.1 mag. From now on, magnitudes in
the SDSS bands are meant to be in the AB system, while for
Johnson-Cousins filters they are expressed in Vega units. Colours that
involve SDSS and Johnson-Cousins pass-bands, are computed by
subtracting the magnitudes in the two respective systems such that,
e.g., $i-H$ means $i_{\mathrm{AB}}-H_{\mathrm{Vega}}$.

\begin{figure*}
\includegraphics[width=\textwidth]{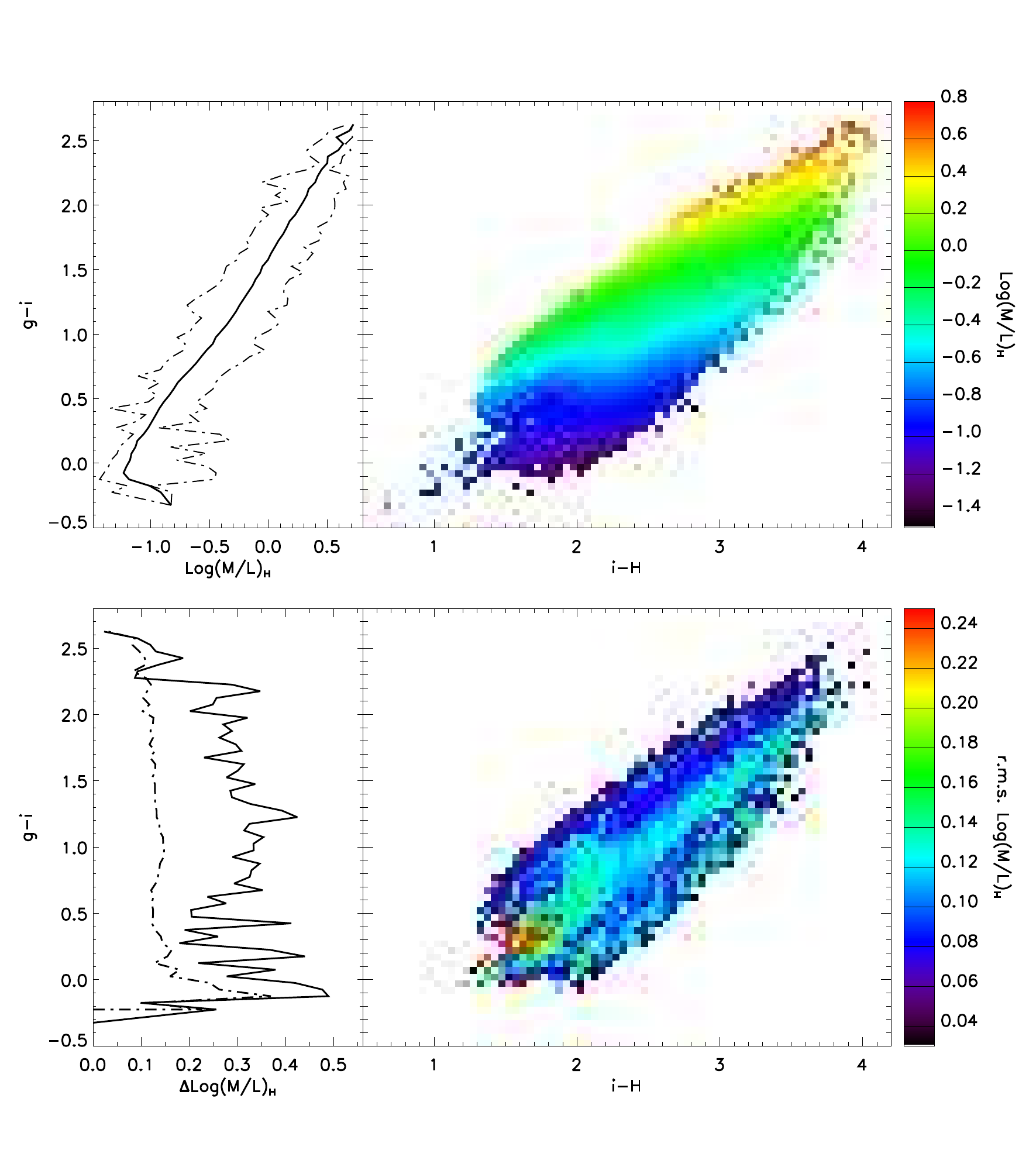}
\caption{Effective mass-to-light ratio in H band as a function of
  colours in the Monte Carlo library based on CB07 SPS models. {\it
    Top panels:} {\it right} the median $\log\Upsilon_H$ for models
  binned $0.05\times0.05$ mag$^2$ in the $(i-H)-(g-i)$ colour-colour
  space. {\it Left}: the median $\log\Upsilon_H$ for the same models
  binned 0.05 mag in $(g-i)$ only (solid line), and the minimum and
  maximum value of $\log\Upsilon_H(i-H,g-i)$ along each row of the
  right panel (dash-dotted lines).  {\it Bottom panels:} {\it right},
  the {\it r.m.s.} of $\log\Upsilon_H$ for models binned
  $0.05\times0.05$ mag$^2$ in the $(i-H)-(g-i)$ colour-colour
  space. {\it Left}: the solid line shows the half-range of values of
  $\log\Upsilon_H$ as a function of $g-i$ from the top right panel,
  compared to the mean {\it r.m.s.} of $\log\Upsilon_H$ in individual
  colour-colour cells (dash-dotted line).}\label{fig_SPS_a}
\end{figure*}

The top right panel of Figure \ref{fig_SPS_a} shows $\Upsilon_H$, the
$M/L$ ratio in $H$-band in solar units, as a function of $(g-i)$,
$(i-H)$ colours, based on the Monte Carlo library of SPS models. Our
models cover a broad sequence from blue (lower left corner) to red
(upper right corner) across the colour-colour diagram, with
$\Upsilon_H$ increasing from $10^{-1.5}$ to $10^{0.8}$, i.e. by a
factor 200. It is apparent that the $g-i$ colour is the main predictor
of the trend in $\Upsilon_H$, which increases by 0.08 dex per 0.1 mag
in $g-i$. However, for a given $g-i$ a range of $\Upsilon_H$ is
allowed. This is quantified in the top left panel of Figure
\ref{fig_SPS_a}, where we plot (solid line) the median $\Upsilon_H$ as
a function of $g-i$ (binned in 0.05 mag). Dash-dotted lines show the
maximum and minimum $\Upsilon_H$ taken from the right-hand panel, for
a given $g-i$: the spread in $\Upsilon_H$ is between 0.5 and 1 dex for
most of $g-i$ values, showing that the additional information from the
second colour, $i-H$, is crucial to minimize the uncertainty in
$\Upsilon_H$.

Even in the 2-dimensional colour space the remaining scatter of the
predicted $\Upsilon_H$ at a given $(i-H)$, $(g-i)$ is significant. The
{\it r.m.s.} of $\log\Upsilon_H$ is represented in the colour-colour
space in the bottom right panel of Figure \ref{fig_SPS_a}. The typical
{\it r.m.s.} ranges between 0.05 and 0.15 dex (i.e., between 10 and 40
per cent approximately) and is thus comparable to the effect of errors
on colours of $\approx 0.1$ mag, except for the region occupied by
``blue'' models ($g-i<0.7 \wedge i-H<2.2$). Models in this region are
characterized by relatively young stellar populations with a strongly
varying NIR emission (in particular by TP-AGB stars, see also Section
\ref{BC03_sect}), which produce {\it r.m.s.} scatter up to 0.25 dex
(approximately a factor 1.8). It is interesting to see how this
scatter in $\log\Upsilon_H(g-i,i-H)$ compares with the half-range of
$\log\Upsilon_H(g-i)$ that we derived from the top right panel of
Fig. \ref{fig_SPS_a}. The latter can be considered an estimate of the
typical error that one makes by replacing $\log\Upsilon_H(g-i,i-H)$ by
the median $\log\Upsilon_H(g-i)$. The bottom left panel of
Fig. \ref{fig_SPS_a} shows that for most $g-i$ the mean {\it r.m.s.}
scatter in $\log\Upsilon_H(g-i,i-H)$ at any given $g-i$ (dash-dotted
line) is $\approx 0.1$ dex, while it is $\approx 0.3$ dex for
$\log\Upsilon_H(g-i)$ (solid line). Through the use of two colours to
determine the $M/L$ ratio one can reduce the uncertainties from a
factor 2 to a factor 1.25.

It is worth noting that the choice of prior parameter distributions in
the model library can affect the estimated median $M/L$ for a given
colour (pair). The time elapsed since the beginning of star formation,
$t_{\rm {form}}$, and the time and intensity of bursts are the
parameters whose distributions affect most the median $M/L$ because
the fraction of mass hidden in old, low-luminosity stars critically
depend on these parameters. As an extreme case, we test the effect of
removing from our library all models with $t_{\rm {form}}<10$~Gyr. As
expected, the models with larger $t_{\rm {form}}$ predict larger
$M/L$. The largest differences (from 0.1 up to 0.4 dex) with respect
to our default library are found for $g-i<0.6$ and for the very
reddest models in $i-H$ at given $g-i$. As it is shown in Section
\ref{sec:cc_space_models_obs} and Fig. \ref{fig_sample_colcol}, these
regions of the colour-colour space are only sparsely populated in the
observations. The rest of the space is only marginally affected
(typically 0.02 dex difference, increasing toward the extreme regions
mentioned before). We conclude that our choice of prior is not
critical when we use two colours. However, it can gain greater
relevance if only one colour is used, as we show in section
\ref{SPS_bell_sec}. In Appendix \ref{physprop_append} we show and
discuss the detailed distributions of the physical parameters that
characterize the models as a function of colours. In particular, we
note that extremely red colours ($g-i > 1.5$) and correspondingly very
high $\Upsilon_H$ (up to 5-6) can be produced only by models with
total dust optical depth $\hat{\tau}_V \gtrsim 3$.

\subsection{Comparison with models using an older TP-AGB star
  prescription (BC03)}\label{BC03_sect}
Most recent works in stellar population synthesis models agree on the
relevance of the TP-AGB phase for a correct estimate of the NIR flux
of stellar populations at ages between 0.3 and 2 Gyr
\citep[e.g.][]{maraston05,bruzual07}. Yet, a fully reliable
quantification of this effect is still under debate. In order to
illustrate the impact of different prescriptions for the TP-AGB phase
(in particular, those concerning its duration) we compare $M/L$
determinations from our standard CB07 models with those derived from
the BC03 models \citep{BC03}, which assume much shorter-lived TP-AGB
stars. We have produced a library of SPS models with identical star
formation history, metallicity, IMF and dust properties as described
before, but using the stellar evolutionary tracks and libraries of
BC03.

\begin{figure}
\includegraphics[width=0.5\textwidth]{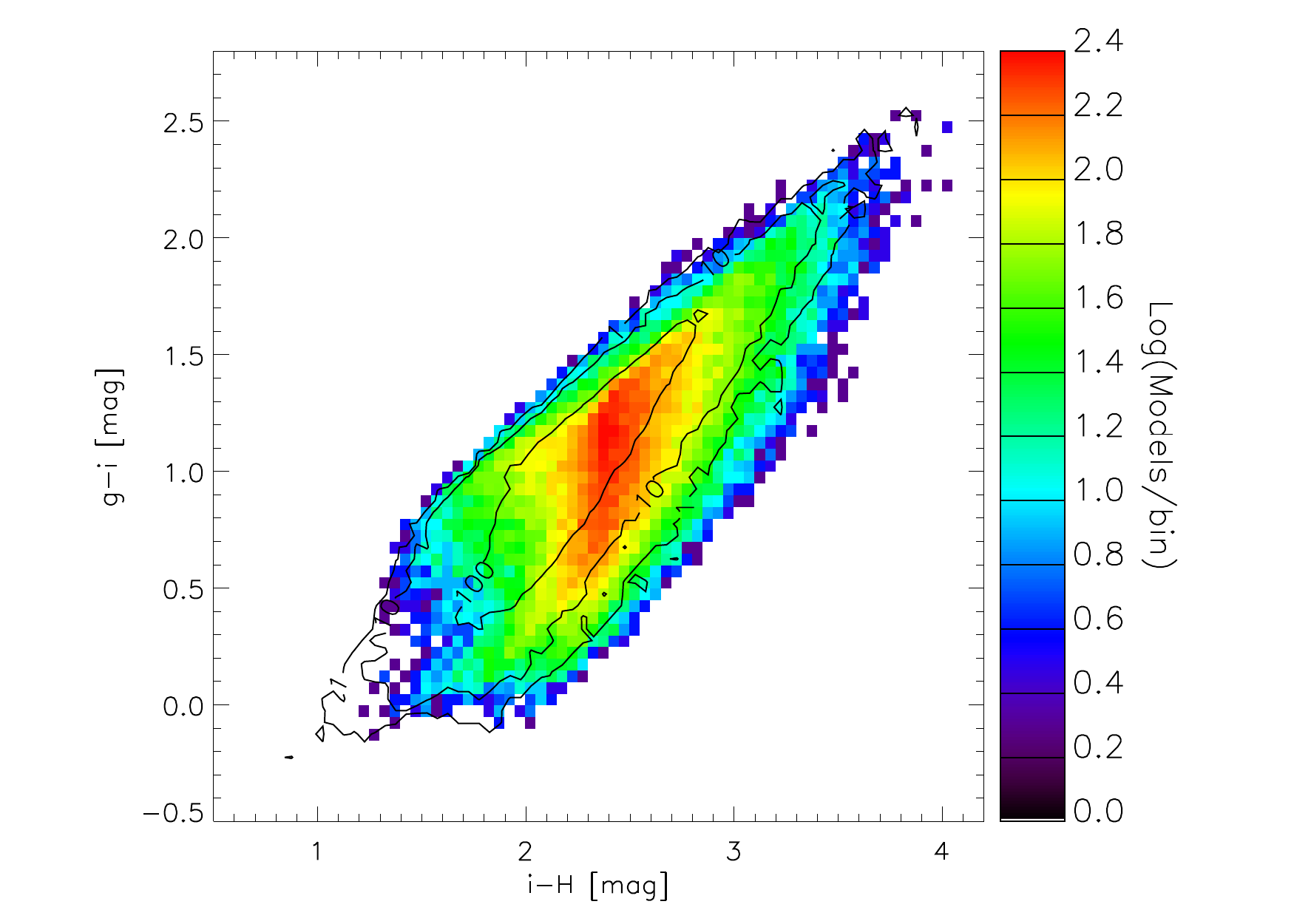}
\caption{Prior distribution of models in $(i-H)-(g-i)$ colour-colour
  space: the colour map shows the distribution for the CB07 library,
  solid contours are for BC03 (1, 10 and 100 models per colour-colour
  bin). The effect of the revised prescription for TP-AGB stars is
  evident in the shift of the CB07 models to redder $i-H$ values with
  respect to BC03.}\label{fig_priors_ac}
\end{figure}

In Figure \ref{fig_priors_ac} we compare the distribution of the two
model libraries in the $(i-H)-(g-i)$ colour-colour space. The
intensity scale in colour shows the distribution of our default
library CB07, while the distribution of BC03 models is shown by the
overlaid contours. The effect of the revised TP-AGB star prescriptions
in CB07 is mainly to move blue models ($i-H<2$, $g-i<1$) to redder
$i-H$. Other regions of the colour-colour space are only slightly
affected by TP-AGB stars, the only significant difference being a
$\sim 0.2$ mag more extended tail at higher $i-H$ for a given $g-i$.

\begin{figure*}
  \includegraphics[width=\textwidth]{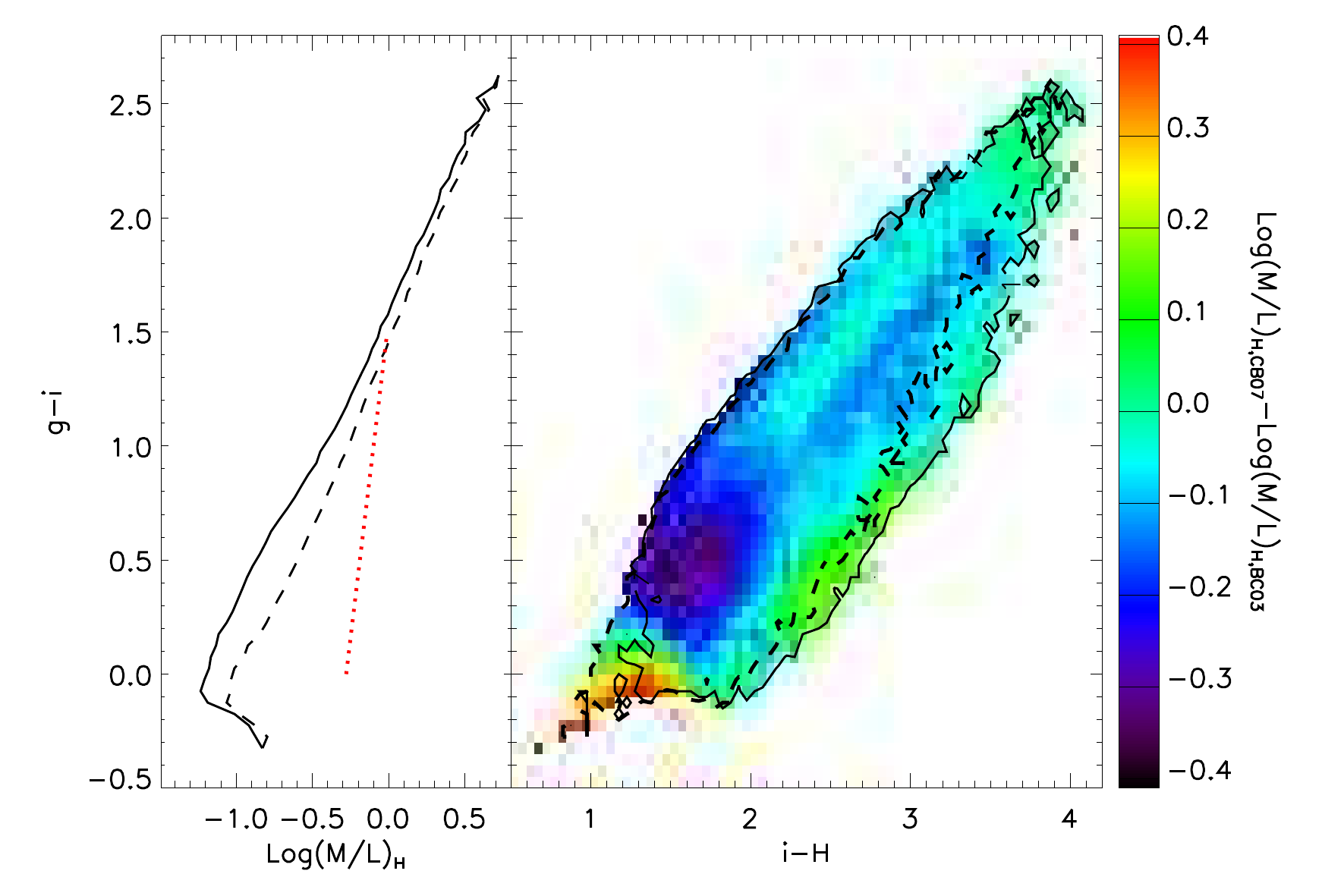}
  \caption{Comparison between CB07- and BC03-based libraries. {\it
      Right panel}: difference of median $\log \Upsilon_H$ derived
    using CB07 and BC03 models, binned $0.05\times0.05$ mag$^2$ in
    $(i-H)-(g-i)$. Black contours show the distribution of models in
    colour-colour space for CB07 (solid contours) and BC03
    (dashed). CB07 models, which include longer-lived TP-AGB stars,
    appear to extend more to red $i-H$ than BC03, with an overall
    shift towards redder $i-H$ at low $g-i$, due to young stellar
    populations which include a significant fraction of TP-AGB. {\it
      Left panel}: median $\log \Upsilon_H$ as a function of $g-i$ for
    CB07- and BC03-based models, shown as solid and dashed lines,
    respectively. The old BC03 models with shorter-lived TP-AGB stars
    overestimate $\Upsilon_H$ by several tenths of dex in blue/young
    stellar populations, with respect to the new CB07. The red dotted
    line represents the power-law fitting formula from Bell et
    al. (2003).}\label{fig_SPS_c}
\end{figure*}

What is most interesting for $M/L$ estimates is the systematic
difference between the two libraries as a function of the observables,
i.e. the colours, which is illustrated in Figure \ref{fig_SPS_c}. In
the right panel different colour shadings shows the logarithmic
difference of the median $\log \Upsilon_H$ between CB07 and BC03. The
original colour space coverage of the two libraries is displayed by
the solid (CB07) and dashed (BC03) contours. Outside the original
coverage (regions where only one of the two libraries has models) the
$M/L$ is extrapolated using a minimum curvature fitting algorithm. As
expected, the largest differences are found for the bluest colours
($g-i<0.7 \wedge i-H<2.2$), that correspond to young ($t<~2$~Gyr)
stellar populations, where the NIR emission is strongly influenced by
TP-AGB stars. At those colours typical systematic differences between
CB07 and BC03 models are between $-0.25$ and $-0.4$ dex, i.e. the old
SPS models that assume short-lived TP-AGB stars imply higher masses,
by a factor $\approx 2$. The rest of the colour space is only slightly
affected, with typical differences between 0 and $-0.1$ dex (20 per
cent). It is worth noting that over all the colour space (with the
only exception of a few marginal regions) CB07 models predict lower
$M/L$ in the NIR than BC03. This systematic offset is also illustrated
in the left panel of Fig. \ref{fig_SPS_c}, where the median
$\log\Upsilon_H(g-i)$ is shown for CB07 and BC03 as solid and dashed
lines, respectively.

\subsection{$M/L$ from optical bands only}
The modeling uncertainties inherent to NIR pass-bands motivated us to
test the accuracy of $M/L$ determination using optical pass-bands and
colours only. For this test we adopt the SDSS $i$ band for luminance
and study the dependence of $\Upsilon_i$ on ($g-i$, $i-H$) (as in the
previous section) and then as a function of $g-i$ only\footnote{We
  have conducted the same test using the SDSS $z$ band instead of $i$
  and obtained qualitatively identical results.}.
\begin{figure*}
\includegraphics[width=\textwidth]{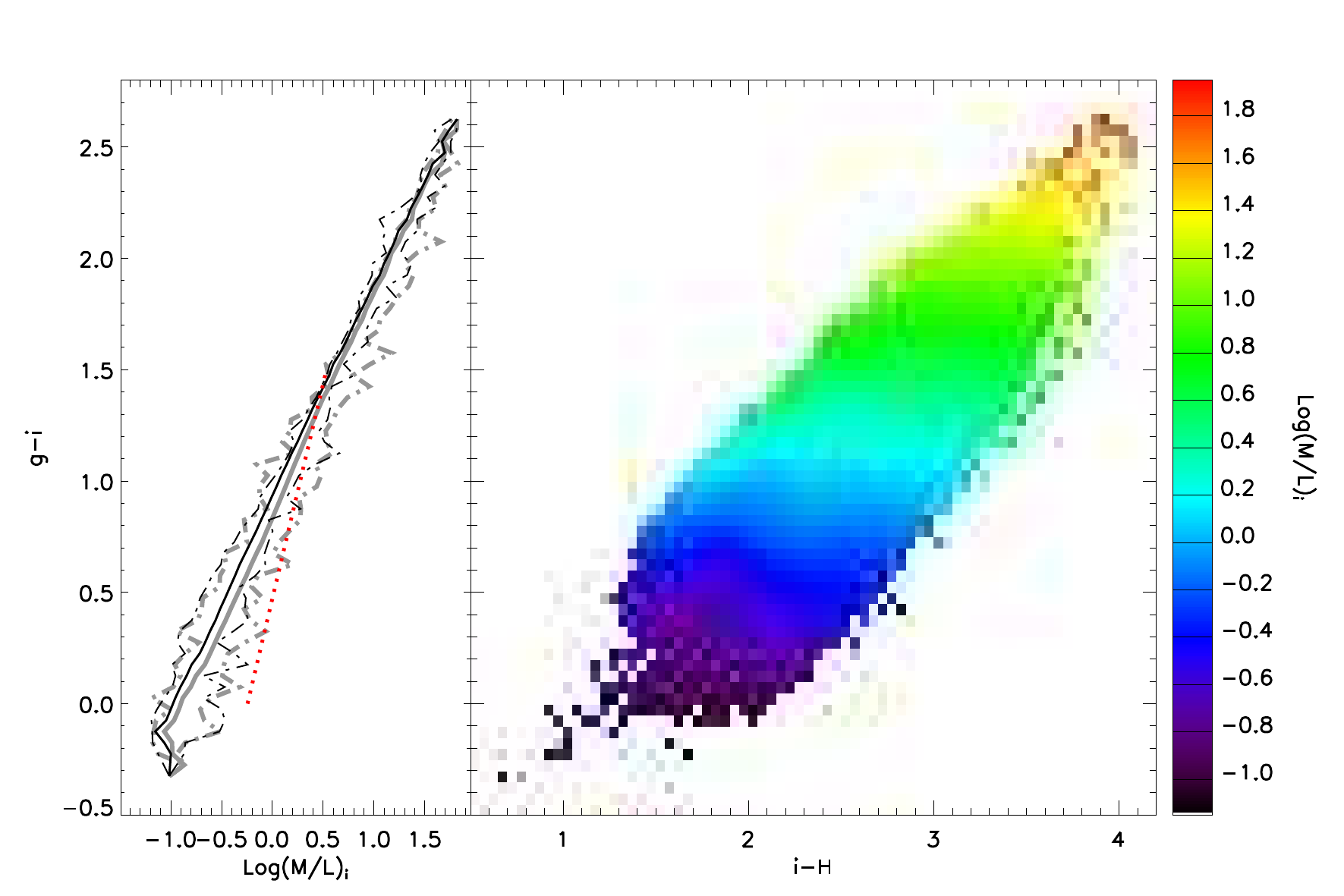}
\caption{Effective mass-to-light ratio in $i$ band as a function of
  colours in the Monte Carlo library based on CB07 SPS models. {\it
    Right panel}: the median $\log\Upsilon_i$ for models binned
  $0.05\times0.05$ mag$^2$ in the $(i-H)-(g-i)$ colour-colour
  space. {\it Left}: the median $\log\Upsilon_i$ for the same models
  binned 0.05 mag in $(g-i)$ (solid line), and the minimum and maximum
  value on $\log\Upsilon_i$ along each row of the right panel
  (dash-dotted lines). Black lines represent CB07 models, while grey
  lines display BC03. The red dotted line is the Bell et al.'s (2003)
  power-law fit.}\label{fig_SPS_d}
\end{figure*}

In the right panel of Figure \ref{fig_SPS_d} we show the effective
$M/L$ in $i$ band as a function of the two optical/NIR colours $g-i$,
$i-H$.  The comparison with the analogous plot of Fig. \ref{fig_SPS_a}
shows two main differences: the range of $M/L$ in $i$ is almost ten
times larger than in $H$ band, but the dependence of $\Upsilon_i$ on
$(i-H)$ is extremely weak. In the left panel we compare the median
$\log\Upsilon_i(g-i)$ (solid black line) to the full range of
$\Upsilon_i$ (black dot-dashed lines) at a given $(g-i)$. The min-max
half range is typically 0.15 dex, comparable to the {\it r.m.s.} in
individual cells in the $(g-i,i-H)$ colour-space that one can derive
from the bottom right panel of Fig.  \ref{fig_SPS_a}.\footnote{The
  {\it r.m.s.}  of $\log\Upsilon_i(g-i,i-H)$ is the same as for
  $\log\Upsilon_H(g-i,i-H)$ because in each cell the luminosity ratio
  in $i$ and $H$ band is fixed by the $i-H$ colour.} In the left panel
of Fig. \ref{fig_SPS_d} we also plot with grey lines median, minimum
and maximum $\log M/L$ as obtained from BC03 models.  The systematic
difference to CB07 models in this case is 0.1 dex at most, to be
compared with differences of up to 0.4 dex in $H$-band (see paragraph
\ref{BC03_sect}).

In the light of these results, the use of $i$-band images for
luminance and $g-i$ maps to extract $\Upsilon_i$ appears as an
attractive alternative to our fiducial method, the use of $H$-band
with $(g-i,~i-H)$ colours. While $H$-band with $(g-i,~i-H)$ is more
accurate and stable against photometric errors because of the smaller
range in $\Upsilon_H$, the use of $i$ and $g$ band is certainly
cheaper in terms of observational resources and less sensitive to the
still controversial modeling of TP-AGB stars. However, in Section
\ref{result_sect} we show that the use of NIR is the only sensible way
to go in two regimes: {\it i)} in presence of heavy dust obscuration,
where radiation at shorter wavelengths will not emerge, and {\it ii)}
in the case of very young stellar populations that completely dominate
the optical light.

\subsection{Comparison with Bell et al. (2003)}\label{SPS_bell_sec}
A one-colour method similar to the one we adopt to derive
$\Upsilon_i(g-i)$ and $\Upsilon_H(g-i)$ was formerly developed by
\cite{bell_dejong01} and subsequently revised by \cite{bell+03}. In
Appendix A2 of their paper they give power-law fits for $M/L$ in
different optical and NIR bands as a function of one optical colour.
In Fig. \ref{fig_SPS_c} and \ref{fig_SPS_d} we plot the Bell et al.'s
relations for $\Upsilon_H(g-i)$ and $\Upsilon_i(g-i)$, respectively,
as red dotted lines, limited to the colour range actually covered in
their work. We have scaled down their $M/L$ by -0.093 dex to take the
difference between our Chabrier IMF and their scaled Salpeter IMF
\citep{gallazzi+08}. Their $M/L$ agree with ours only for the reddest
$(g-i)$ corresponding to old unextincted stellar population. Yet, the
dependence on $(g-i)$ is much weaker according to Bell et al.'s fits
and results in large discrepancies at the blue end of the
distribution, where they predict $M/L$ approximately ten times larger
than ours.  The same systematic variations occur in all bands and
colours analyzed in \cite{bell+03}, as can be verified using the
power-law fits derived from our fiducial models which are given in
Table \ref{plaw_fits_tab} in Appendix \ref{plaw_fits_append}.

There are in fact a number of differences between our methodology and
Bell et al.'s. First of all, their models are based on BC03 SPS
models. However, Figures \ref{fig_SPS_c} and \ref{fig_SPS_d} show that
large discrepancies are present also in comparison to our BC03-based
estimates. \cite{bell+03} do not explicitly take dust into account
and, more importantly, only consider relatively smooth star formation
histories, starting 12 Gyr in the past and with a maximum stellar mass
contribution from burst in the last 2 Gyr smaller than 10 per cent. As
opposed, we do model the effect of dust up to optical depth of $\hat
\tau_V\approx 6$ and, more importantly, in our library we allow young
ages (of a few Gyr) and have a large fraction of star formation
bursts, which are the main cause of our lower $M/L$ for young stellar
populations \citep[see e.g. Fig. 5 of][and Sec. \ref{SPS_sec}
above]{bell_dejong01}, especially when one colour only is used. On the
local scales that we want to study, both dust and bursty star
formation histories cannot be neglected. This is easy to realize just
looking at the true colour images of common spiral galaxies, where
dust reddened regions are seen and young OB associations dominate in
spiral arms. Therefore we argue that our models are better suited to
describe SEDs on local scales than Bell et al.'s.  This may not be the
case if galaxies are considered globally: in fact for ``normal''
galaxies, global star formation histories are likely to be much
smoother and dust is not expected to play a major role (except for
edge-on disks). We explore this issue further in Section
\ref{mass_bias_sec}.

\section{Sample and imaging data}\label{sample_sec}
To test our mass map reconstruction we select a small sample of nearby
galaxies that span a broad range of morphologies and physical
properties and for which a wealth of high-quality multi-wavelength
imaging is available. We draw our sample from the Spitzer Infrared
Nearby Galaxies Survey \citep[SINGS][]{SINGS_kennicutt+03}, a
comprehensive imaging and spectroscopic study of 75 nearby galaxies (D
$< 30$ Mpc) conducted in the IR with the Spitzer Space Telescope, for
which coordinated observations at visible, near-IR, ultraviolet, and
radio wavelengths are either already in place or planned.
Complementing this unique dataset with high-quality stellar mass maps
will provide key insights into the physics of galaxies and, at the
same time, will allow us to test our method in the best characterized
physical conditions.  Among the 75 SINGS galaxies we select 9 for
which SDSS and medium/deep near-IR (1-2.5 $\mu$m) imaging is
available. The latter is taken either from GOLD Mine \citep{goldmine},
a large database that provides NIR images of 1568 galaxies (mainly in
the Virgo cluster and the Coma super-cluster), or from the 3rd data
release of UKIDSS \citep[][Warren et al., in
preparation]{lawrence+07}, that uses the UKIRT Wide Field Camera
\citep[WFCAM][]{casali+07}.

The 9 galaxies are listed in Table \ref{sample_tab} with their NGC
name (column 1), coordinates (col. 2 and 3) and morphological type
(col. 4) according to the RC3 catalog \citep{RC3}. Distances are
reported in column 5. Seven out of the nine galaxies belong to the
Virgo cluster and, following \cite{gavazzi+99}, we assign them a
distance of 17.1 Mpc. For the other two galaxies we use distances as
given by NED based on their measured redshift, assuming $H_0=73~
\mathrm{km~sec^{-1}~Mpc^{-1}}$. Based on the computed distance and the
pixel scale of the different detectors we compute also the angular
scale (col. 6) and pixel scale (col. 7): for the median seeing of
$\approx 1.4$'', we could resolve physical scales of $\approx 120$ pc
in all cases, although the pixel scales range from 18 to 132 pc. We
use $H$ as NIR band except for NGC\,4569, for which only $K_s$ is
available.
\begin{table*}
\begin{minipage}{\textwidth}
\caption{The sample}\label{sample_tab}
\begin{tabular}{lrrcrrrcc}
  \hline
  Denomination & RA & Dec  & Morph. type & Distance & Angular scale & Pixel scale & NIR & NIR source \\
               & (J2000.0)  & (J2000.0) & & Mpc & pc arcsec$^{-1}$ & pc pixel$^{-1}$       \\
  (1) & (2) & (3)  & (4) & (5) & (6) & (7) & (8) & (9) \\
  \hline

  NGC\,3521 & 11h05m48.6s & $-$00d02m09s  & SABbc   & 9.2 &  45 &  18 & H & UKIDSS \\
  NGC\,4254 & 12h18m49.6s & +14d24m59s    & SAc     &17.1 &  82 & 132 & H & GOLDMine \\
  NGC\,4321 & 12h22m54.9s & +15d49m21s    & SABbc   &17.1 &  82 & 132 & H & GOLDMine \\
  NGC\,4450 & 12h28m29.6s & +17d05m06s    & SAab    &17.1 &  82 & 124 & H & GOLDMine \\
  NGC\,4536 & 12h34m27.0s & +02d11m17s    & SABbc   &17.1 &  82 & 132 & H & GOLDMine \\
  NGC\,4552 & 12h35m39.8s & +12d33m23s    & E       &17.1 &  82 &  33 & H & UKIDSS \\
  NGC\,4569 & 12h36m49.8s & +13d09m46s    & SABab   &17.1 &  82 & 132 & K$_s$ & GOLDMine \\
  NGC\,4579 & 12h37m43.5s & +11d49m05s    & SABb    &17.1 &  82 & 132 & H & GOLDMine \\
  NGC\,5713 & 14h40m11.5s & $-$00d17m20s  & SABbc p &25.9 & 126 &  51 & H & UKIDSS
\end{tabular}
\end{minipage}
\end{table*}
Our sample includes one elliptical galaxy, two early type Sab spirals,
one Sb, four Sbc's (including one peculiar) and one Sc, thus spanning
the whole range of morphologies for ``normal'' (i.e. not irregular)
galaxies.

\begin{figure*}
\includegraphics[width=\textwidth]{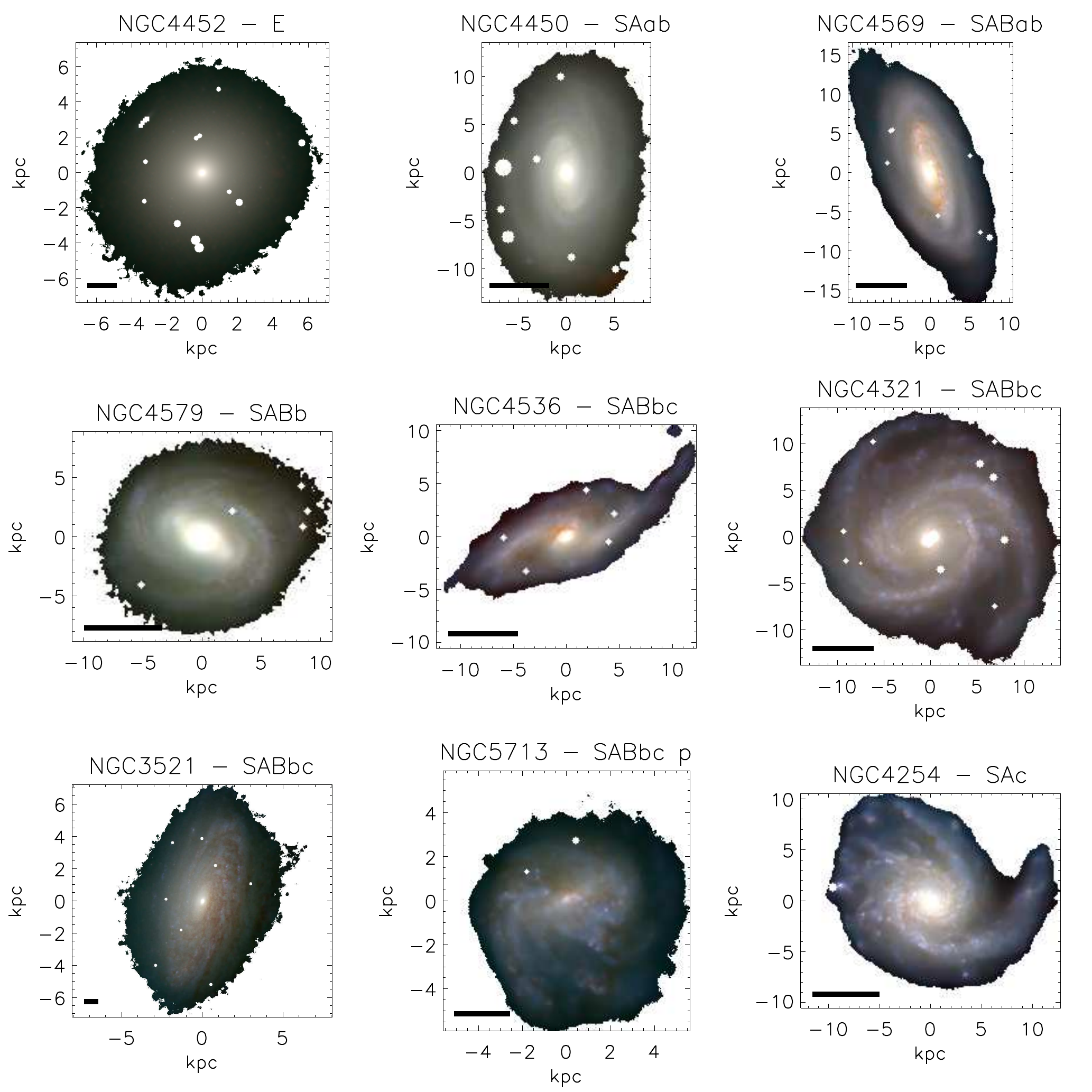}
\caption{$H$ (red channel) - $i$ (green) - $g$ (blue) colour composite
  image of the nine galaxies in the sample, ordered according to their
  Hubble type. Images are adaptively smoothed with {\sc adaptsmooth}
  to ensure a minimum signal-to-noise ratio of 20 at each pixel and
  foreground stars are masked out, as explained in Section
  \ref{imaging_sec}. Scales are in physical kpc. The black rod in each
  panel corresponds to 50 pixels.}\label{fig_sample_RGB}
\end{figure*}

True colour images of the galaxies sorted by morphological type are
presented in Figure \ref{fig_sample_RGB}. The NIR band is mapped in
the red channel, $i$ band in the green and $g$ in the blue; the three
channels are shown in logarithmic intensity scaling and are balanced
to show a solar spectral energy distribution as white. The images that
we show in Fig. \ref{fig_sample_RGB} are matched, calibrated and
filtered as explained in Section \ref{imaging_sec}. Each panel reports
the physical scale in kpc and a rod whose length corresponds to 50
pixels.
\subsection{Image reduction and calibration}
Imaging data for this study are taken from three different sources,
each requiring slightly different pre-reductions and calibrations,
that we describe in this section.

For the optical images ($g$, $i$ and $z$ bands) we completely rely on
the 7th data release of the SDSS \citep[][ Abazajian et al. 2009, in
press]{SDSS}. SDSS images come in the format of ``corrected frames'',
that are bias-subtracted, flat-fielded cuts of long imaging
scans. Given the relatively large size of our galaxies, in some cases
two different scans must be combined. In the general case we reduce
each scan separately: first we join the frames, then we compute an
accurate astrometric solution using stars from the SDSS catalog. We
subtract the sky background by fitting a plane surface to the pixels
in a series of boxes that we define around the galaxy, with a typical
size of roughly one tenth of the galaxy. The fitted plane is allowed
to be tilted in the scan direction only, in order to take temporal
background variations into account. The {\it r.m.s.} of the median
background levels among the boxes provide an estimate of the large
scale background fluctuations that is used later on to make S/N
cuts. If more than one scan is used, we choose a primary scan (where
most of the galaxy is contained) and we rescale all secondary scans in
intensity according to the difference in photometric zero point (as
derived from the SDSS database). Finally, we build a mosaic of the
scans based on the previously computed astrometric solutions using
{\sc swarp} ({\tt
  http://astromatic.iap.fr/software/swarp}).\footnote{The automated
  procedure to build SDSS mosaics as described here is available as
  {\sc iraf} package at {\tt
    http://www.mpia.de/homes/zibetti/software/SDSSmosaic.html}} We
refer the reader to
\cite{fukugita_etal96,gunn_etal98,gunn+06,smith_etal02} for a summary
of the properties of the SDSS imaging.

The NIR images from GOLD Mine come as reduced and calibrated coadded
images. We compute accurate astrometric solutions for each image by
matching $i$-band selected stars from the SDSS catalogs. Sky
subtraction is performed similarly as for SDSS images, but using a
constant plane (order 0 surface in both coordinates). Finally we check
(and refine where needed) the photometric calibration using stars from
2MASS \citep{2MASS}: the final photometric accuracy (zero point) is
typically $\lesssim 0.1$~mag. We note that all of the GOLD Mine images
used in this work have pixel scales between 1.5 and 1.6 arcsec per
pixel, that severely under-sample the PSF. This does not represent a
problem for the following analysis though, as a sufficient physical
resolution is provided anyway (Table \ref{sample_tab}).

For UKIDSS images we have used stacks from the 3rd data release, which
is described in detail in Warren et al. (in preparation). The pipeline
processing and science archive are described in Irwin et al.  (2009,
in preparation) and \cite{hambly+08}. Sky subtraction and astrometric
calibrations are performed exactly as for GOLD Mine images. The UKIDSS
photometric system is described in \cite{hewett+06}, and the
calibration is described in \cite{hodgkin+08}.  The absolute
photometric accuracy is typically around few 0.01 mag. We note that
the pixel scale of 0.4 arcsec per pixel perfectly matches SDSS images.

In both GOLD Mine and UKIDSS datasets the typical depth reached by the
NIR images used in this study is $\mu_H\approx
20.5~\mathrm{mag~arcsec^{-2}}$ ($3~\sigma$ on a square arcsecond).

All fluxes are corrected for Galactic foreground extinction, as given
in NED or the SDSS database, which are based on \cite{schlegel_dust}.

\section{Methodology (II): from multi-band images to mass
  maps}\label{imaging_sec}
\subsection{Image processing}
Images in different bands must be registered and resampled to a common
resolution before pixel-by-pixel colour information can be
extracted. To do this we use {\sc Swarp} and the astrometric solutions
computed in the previous section. We choose to degrade all
sky-subtracted images to the lowest resolution image for each
galaxy. In practice, for all images taken from GOLD Mine this
translates into degrading the SDSS images to the NIR pixel
scale\footnote{In order to correctly propagate the noise properties of
  the images while degrading the resolution it is important to run
  {\sc Swarp} with an oversampling factor equal to the ratio between
  the final and the original pixel scale.}, while the original pixel
scale is kept for the UKIDSS data. In principle, images taken with
different instruments, in different bands and seeing conditions must
also be convolved to a common point spread function (PSF). However we
do not apply such convolution, as the PSFs are already similar and a
convolution would corrupt the noise properties.

As shown in Sec. \ref{SPS_sec}, colours must be accurate at better to
$\lesssim 0.1$~mag to compute $M/L$ ratios that match theoretical
uncertainties. In turn, this requires surface brightness in each band
to be accurate within $\approx 0.05$ mag, or $S/N\gtrsim 20$ per
pixel. The noise budget includes local photon noise, assumed to be
Gaussian, and background fluctuations that may become the dominant
source of uncertainty at low surface brightness.

Local photon noise can be reduced to the required level with low-pass
filters, using a smoothing kernel of sufficiently large size. However,
a fixed-width kernel produces a uniform degradation of the effective
spatial resolution of the entire image, including bright regions of
the galaxy where no or minimal smoothing is required. For this reason
we have implemented a new code to perform image smoothing with a
variable kernel, whose size is adapted to the local $S/N$. This code
is called {\sc adaptsmooth} and will be presented in detail in a
forthcoming paper (Zibetti 2009, in preparation). Briefly, the idea is
to replace the intensity in each pixel with the median intensity in a
circle of radius $R$ of surrounding pixels, where $R$ is determined as
the minimum radius required to attain the minimum $S/N$ of 20. The
procedure works by increasing $R$ iteratively. If the minimum $S/N$ of
20 cannot be reached even with the maximum smoothing radius
$R_{\mathrm{max}}$, the pixel is flagged and assigned a value of 0. In
this work we adopt $R_{\mathrm{max}}=13$ pixels for the GOLD Mine
images (corresponding to 20 arcsec) and 20 for the UKIDSS
(corresponding to 8 arcsec). In this way the full spatial resolution
is preserved in the brightest regions of a galaxy, while increasingly
strong smoothing is applied to lower and lower surface brightness
regions.

{\sc adaptsmooth} is run a first time on the individual images in each
band. A mask that contains the smoothing radius for each pixel (or an
overflow value where the required $S/N$ cannot be reached within
$R_{\mathrm{max}}$) is output for each image. In order to match the
spatial resolution between all three bands we combine the masks into a
common mask with the maximum of the three smoothing radii at a given
position. We then apply an intensity cut to take into account large
scale background fluctuations, as computed from the sky box
statistics. All pixels with an intensity in the smoothed image less
than 10 times the large scale background fluctuation {\it r.m.s.} in
one of the three bands are flagged with the overflow value in the
mask. Furthermore we manually edit the mask to flag stars and other
interlopers. With this mask we re-run {\sc adaptsmooth} in all bands
in ``input mask'' mode, that is using the smoothing radii as given in
the input mask. In pixels where the overflow value is set, a default
value for undefined is output. The adaptively smoothed images that
result from this procedure are shown in the three-colour composite
images of Figure \ref{fig_sample_RGB}. We note that {\it i)} the NIR
images put the strictest constraints for smoothing and intensity cuts;
{\it ii)} the intensity cut we adopt here is less strict than required
to ensure $S/N>20$ at all positions and can produce systematic colour
offsets. However, the intensity threshold in the NIR in all cases is
so high that background fluctuations in the optical are negligible in
the regions that make it through the cut. Effectively, the intensity
cut ensure that the error on colour due to background fluctuations is
roughly the same as the error on the $H$-band only, that is 10 per
cent at most.

From the adaptively smoothed and matched images in the three (or two)
bands we compute colours and surface brightness in solar units per
pc$^2$ (in the ``luminance'' band), for each pixel. From colours we
derive the $M/L$ ratio as explained in Sec. \ref{SPS_sec} and can
multiply it by the surface brightness in order to obtain the stellar
mass surface density.

\subsection{Models and observations in the colour-colour
  space}\label{sec:cc_space_models_obs}
Before illustrating the results of our stellar mass map reconstruction
method, we check to which extent the models can actually reproduce the
observed pixel-by-pixel colours in galaxies.
\begin{figure*}
\includegraphics[width=\textwidth]{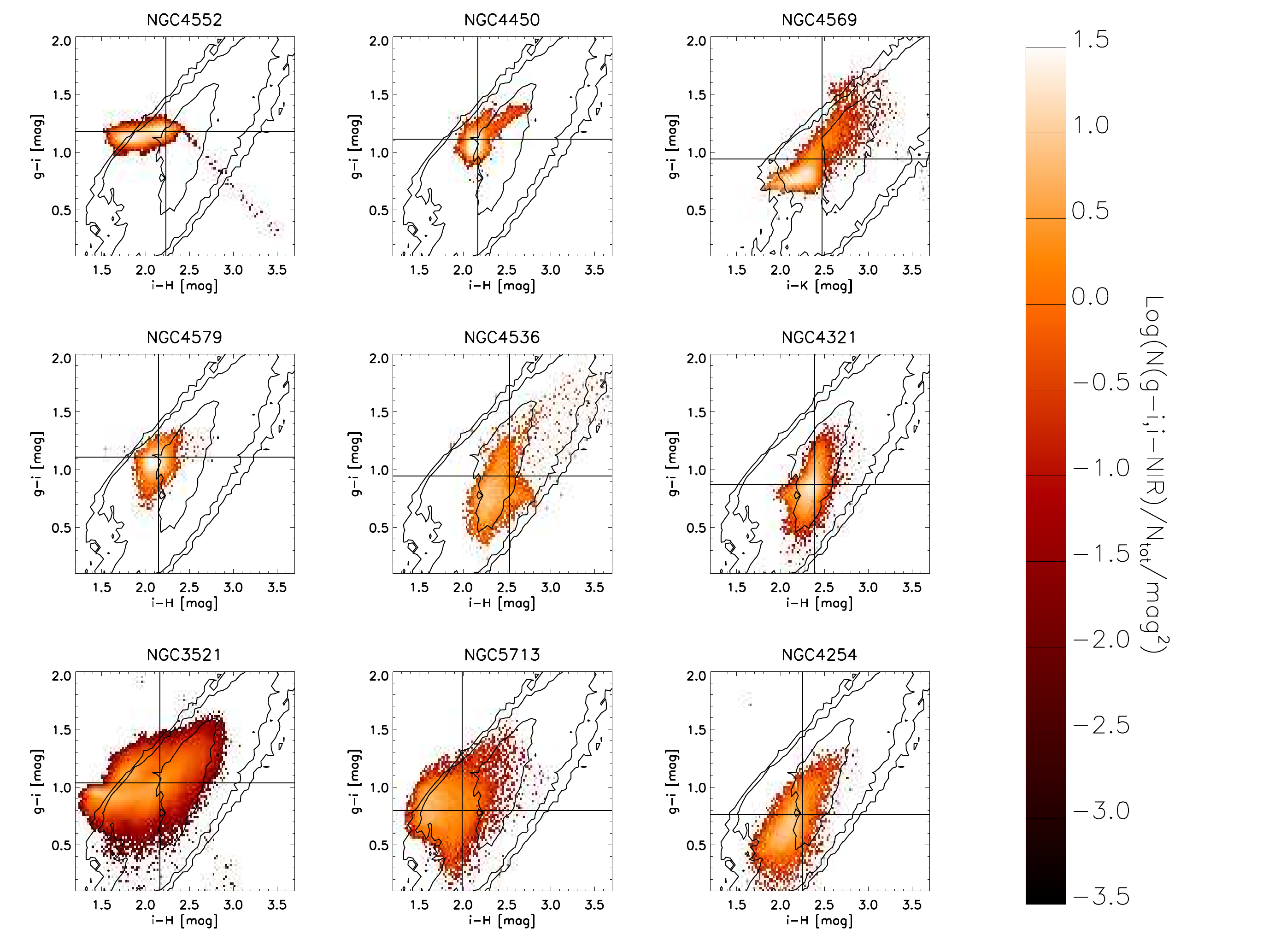}
\caption{Distribution of pixels in the $(i-H)$-$(g-i)$ colour-colour
  space, given in fraction of pixels per cell of mag$^2$. Black
  contours represent the prior distribution of models (1, 10 and 100
  models per bin) in the CB07 library. The models match the observed
  distributions well, with a few minor exceptions discussed in the
  text. }\label{fig_sample_colcol}
\end{figure*}
In Figure \ref{fig_sample_colcol} we plot the distribution of pixels
in the $(g-i,i-H[K_s])$ space for the nine galaxies in our sample. The
relative number density of pixels per colour cell is displayed
according to the colour key on the right side. On the top of it we
overlay contours showing the number density of models in our library.
We observe that the vast majority of pixels in our galaxies lie within
the contours that trace the distribution of models. Although this is
{\it not} a proof that models are correct, it reassures us that we can
reproduce the observations. On the other hand, a comparison with
Figure \ref{fig_physprop} shows that broad ranges in all physical
parameters are required for the models to match the observed colours.

Few pixels have colours are not covered by models, in NGC\,4552,
NGC\,3521 and NGC\,5713. Their $M/L$ ratio must be derived via
extrapolation (minimum surface curvature fitting). We have scrutinized
these pixels that lie blue-ward of the model contours (in $i-H$) and
found that they are from the lowest surface brightness regions. A
possible explanation for this could be just the influence of
background fluctuations in the NIR, that can certainly account for
errors of $\approx 0.1$~mag. Further, in NGC\,4552 the data shows at
face-value a strong $i-H$ colour gradient, with bluer values at larger
radii, which does not correspond to any similar trend in $g-i$. A
non-uniform background cannot explain this effect, since the strength
of this gradient appears the same at different position angles. A
metallicity gradient can also be invoked, but not as a full
explanation, since we do have low metallicity models in our library
and yet we are unable to recover such blue $i-H$. \cite{michard02} and
\cite{wu+05} have pointed out in the past that optical thinned CCDs
can have very extended PSF wings (especially in $i$ band), up to
arcminute scales. Such wings are not expected in the NIR, although no
studies of the phenomenon have been conducted so far. We can speculate
that large-angle PSF wings from scattering in $i$ (and $g$, but not in
$H$) band cause the ``blue'' $i-H$ halo around NGC\,4552 (and possibly
the blue ``halo'' around NGC\,3521 and 5713). Indeed, we observed the
very same effect also in another elliptical of similar apparent size
and luminosity, NGC\,4621 (which is not in the current sample).

Despite those possible systematics at low surface brightness levels,
Fig. \ref{fig_sample_colcol} suggests that our method is on a solid
footing for regions within the classically defined optical radius of a
galaxy.

\section{Results}\label{result_sect}
\subsection{$M/L$ and mass maps}\label{massmaps_subsect}
\begin{figure*}
\includegraphics[width=\textwidth]{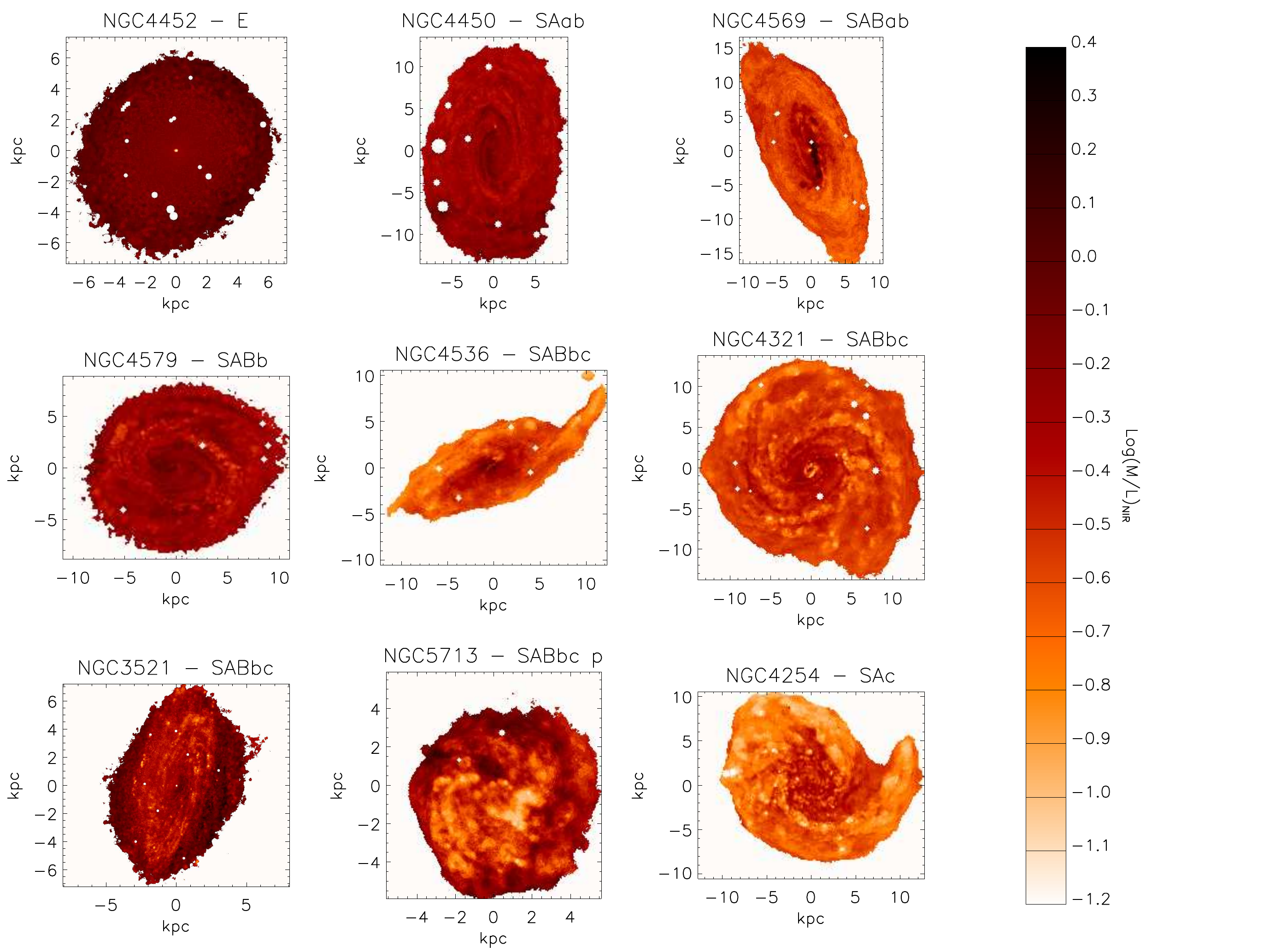}
\caption{Maps of $\log\Upsilon_H$ for the nine
  galaxies.}\label{fig_sample_ML}
\end{figure*}

In Figure \ref{fig_sample_ML} we show the resulting $M/L$ maps
($H[K_s]$-band) for the nine galaxies, with light (dark) tints
representing low (high) $M/L$, as indicated in the side colour
key. 
Early type galaxies tend to have uniform $M/L$, due to their
phase-mixed stellar populations and lack of substantial dust
obscuration\footnote{Note the bright nucleus in NGC\,4452: the LINER
  shines in blue $g-i$ colour that is ``interpreted'' by our algorithm
  as young stellar populations. The nuclear pixels can also be seen as
  a track directed towards the lower-right corner in
  Fig. \ref{fig_sample_colcol}.}. In later type galaxies young, blue
stellar populations in the spiral arms result in a spiral structure of
lower $M/L$. For the two grand-design spirals (NGC\,4321 and 4254) the
radial decrease of $M/L$ is a clear effect of the younger, lower
metallicity stars that populate the outer disk \citep[see
e.g.][]{portinari_salucci09}. The presence of an old/metal rich bulge
is at the origin of the high-$M/L$ regions in the inner parts of most
spirals, except NGC\,5713 which has a peculiar morphology. Dust lanes,
which are observable in the true colour images of
Fig. \ref{fig_sample_RGB} as reddish intrusions, are highlighted in
the $M/L$ maps by the most extreme high values, as one expects as a
consequence of light absorption.

\begin{figure*}
\includegraphics[width=\textwidth]{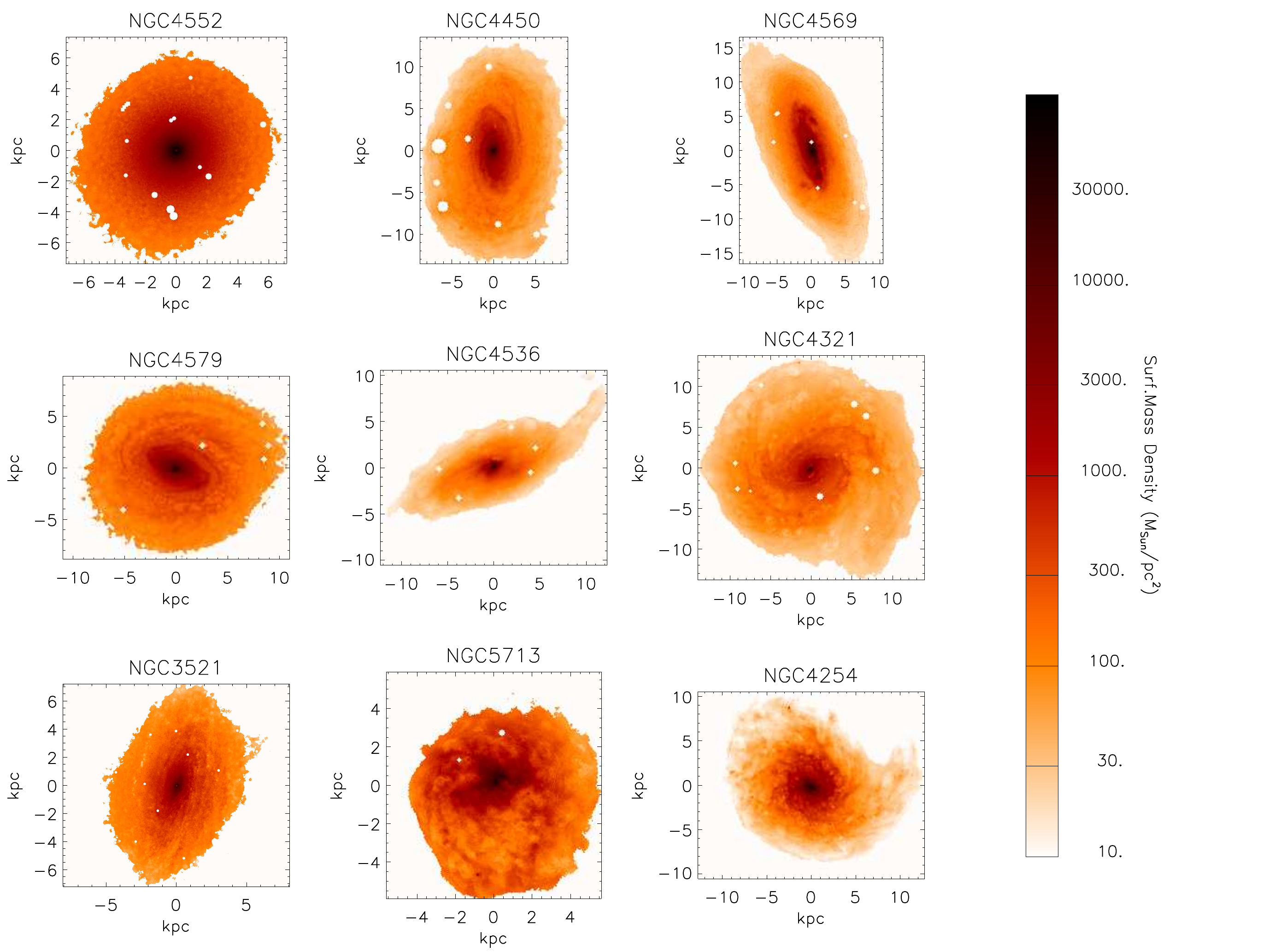}
\caption{Stellar mass surface density maps for the nine
  galaxies.}\label{fig_sample_mass}
\end{figure*}
The stellar mass maps, resulting from multiplying $H[K_s]$-band
intensity with the $M/L$ of Fig. \ref{fig_sample_ML}, are shown in
Figure \ref{fig_sample_mass}. What is most striking here is the
overall smoothness of the stellar mass distribution across the entire
morphological sequence. The prominent spiral arms which are seen in
the true colour images (and in the individual bands) are greatly
reduced in the mass maps. We will quantitatively analyze the relative
bias of galaxy structure in mass maps versus brightness maps at
different wavelength in paper II of this series. To give a more
quantitative idea of how structure changes from a light-weighted to a
mass weighted view, we note that the relative arm-interarm contrast
for NGC\,4321 decreases roughly by a factor 2 when we measure it from
the mass map rather than from $i$ or $H$ band images.

\subsection{Comparison between different methods}\label{sec_map_comp}
\begin{figure*}
\includegraphics[width=\textwidth]{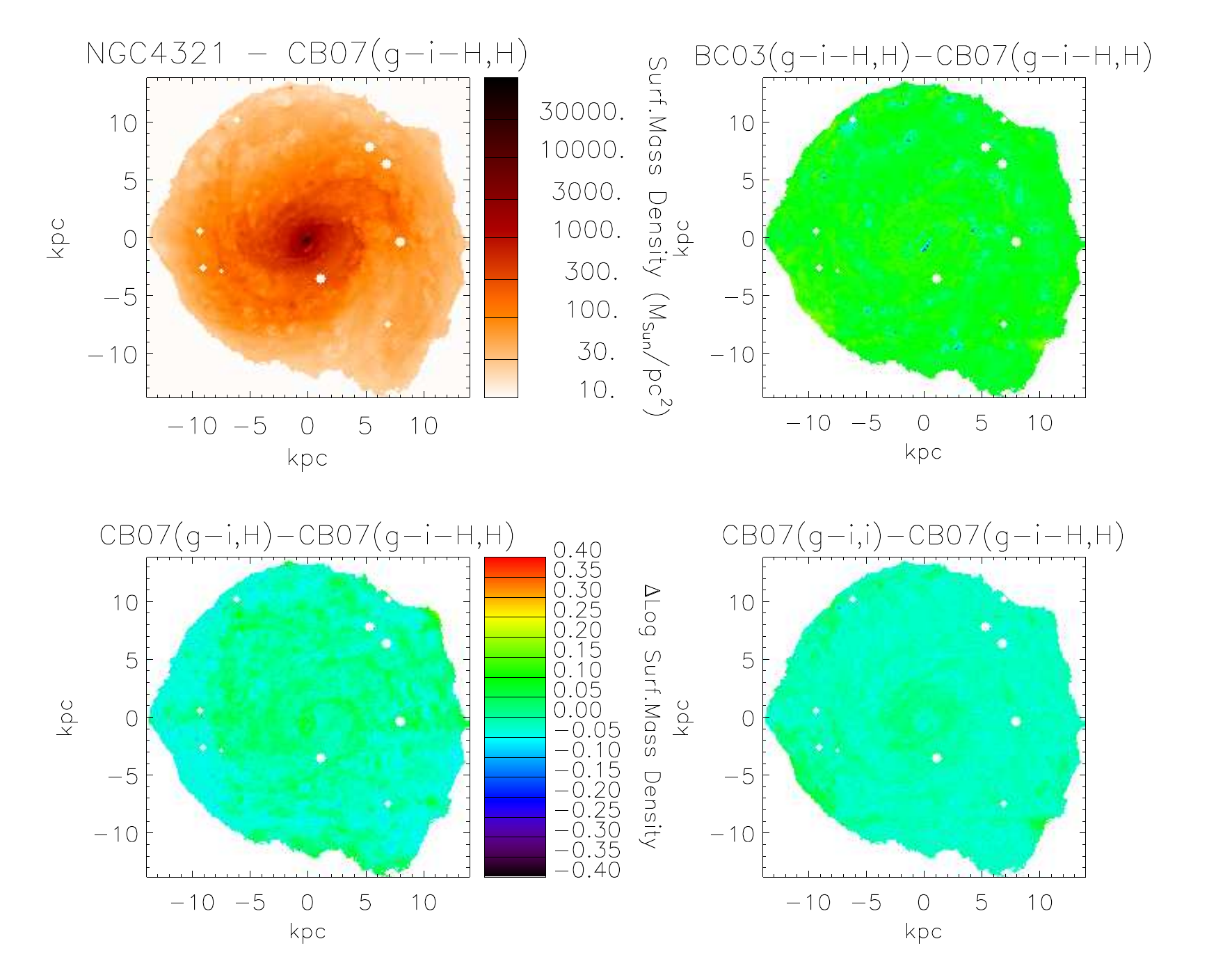}
\caption{Comparison between stellar mass maps obtained with different
  methods/models for NGC4321 (M100). {\it Top left panel} is the map
  obtained with our fiducial model CB07 using two colours $(i-H)$,
  $(g-i)$ to extract $M/L$ in H-band. The other three panels show the
  logarithmic difference between three alternative methods and the
  fiducial one. In the {\it top right panel} we use the same method
  applied to BC03 models. In the {\it bottom left panel} we use CB07
  again, but $\Upsilon_H$ is derived from $g-i$ alone (see
  Fig. \ref{fig_SPS_a}, top left panel). Finally, in the {\it bottom
    right panel} $M/L$ is determined in $i$ band from $g-i$ alone (see
  Fig. \ref{fig_SPS_d} left panel) from CB07
  models.}\label{fig_N4321_mass_comp}
\end{figure*}

\begin{figure*}
\includegraphics[width=\textwidth]{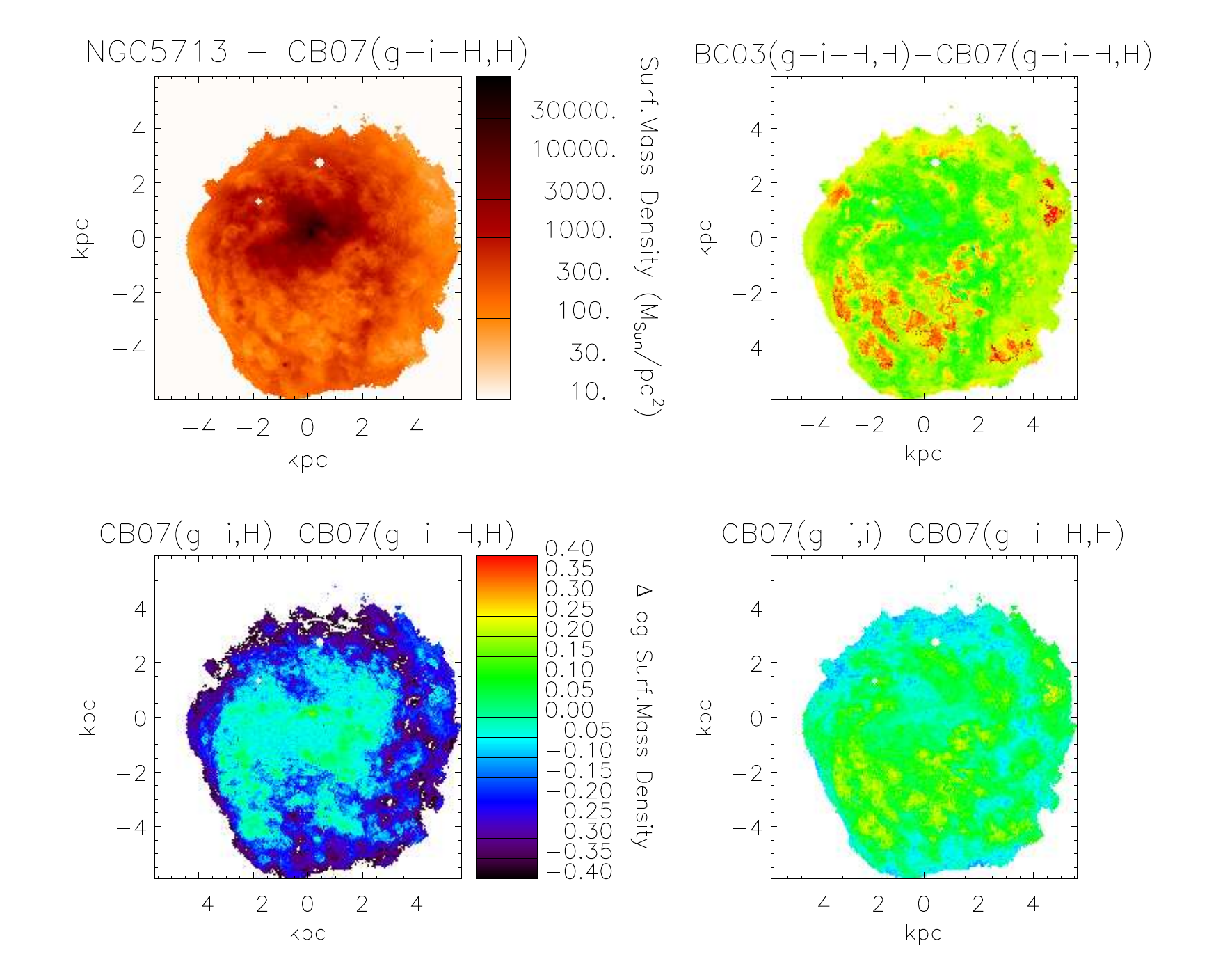}
\caption{Same as Fig. \ref{fig_N4321_mass_comp} but for
  NGC5713.}\label{fig_N5713_mass_comp}
\end{figure*}

We now compare the stellar mass maps obtained with the different
methods described in paragraph \ref{SPS_sec}. We consider (a) our
fiducial method based on $\Upsilon_H(i-H,g-i)$ as derived from CB07
SPS models, in conjunction with $H$-band images; (b)
$\Upsilon_H(i-H,g-i)$ from BC03 SPS models, with $H$-band images; (c)
$\Upsilon_H(g-i)$ from CB07 SPS models, with $H$-band images; (d)
$\Upsilon_i(g-i)$ from CB07 SPS models, in conjunction with the
$i$-band images. In particular, we present such comparisons for
NGC\,4321, which is representative of normal galaxies with minor dusty
regions and moderate star formation activity, and for NGC\,5713, which
is the extreme case in terms of dusty and intensely star forming
regions. Mass maps comparisons for NGC\,4321 and 5713 are presented in
Figures \ref{fig_N4321_mass_comp} and \ref{fig_N5713_mass_comp},
respectively. In both figures, the top left panel shows the stellar
mass map obtained with the fiducial method (a). The other three panels
display the logarithmic difference between the mass maps obtained with
methods (b), (c) and (d), respectively, and that from the fiducial
method (a).  For NGC\,4321 we observe that the four methods result in
very similar structure, with {\it r.m.s.}  in the residuals between 5
and 7 per cent. Method (d) ($\Upsilon_i(g-i)$ from CB07) provides the
closest match to the default method (a), with an average
pixel-by-pixel offset of 3 per cent and residual {\it r.m.s.} of 5 per
cent. Using $g-i$ alone to constrain $M/L$ in $H$ band results in
worse agreement with method (a) ($r.m.s.=7$~per cent). This is
expected from Fig. \ref{fig_SPS_a} and \ref{fig_SPS_d}: while
$\Upsilon_i$ shows little dependence on $i-H$, on the contrary
$\Upsilon_H$ does significantly depend on both colours. As for the
difference between using CB07 or BC03 (method a and b), we note a
systematic offset of approximately $+0.1$ dex going from CB07 to BC03,
although the difference map looks very uniform.

Contrary to the ``normal'' galaxy NGC\,4321, NGC\,5713 is
characterized by intensely star forming regions and prominent dust
lanes. Models based on CB07 and BC03 predict very different $M/L$
especially in presence of very young stellar populations (see
Fig. \ref{fig_SPS_c}), as shown in the top-right panel of
Fig. \ref{fig_N5713_mass_comp}. All blue regions
(cf. Fig. \ref{fig_sample_RGB}) have masses over-estimated by up to
0.4 dex (2.5 times) in BC03 models.
A qualitatively similar (but quantitatively smaller, up to 0.25 dex
only) over-estimate of the mass of regions dominated by young stellar
populations arises from method (d): in this case the median
$\Upsilon_i(g-i)$ is not representative for these extreme stellar
populations that lie to the leftmost edge of the model colour
distribution (see Fig. \ref{fig_sample_colcol}). This appears to be
the case also for method (c) and demonstrates the need for a second
colour to properly describe this region of the parameter space.

\begin{figure}
\includegraphics[width=0.5\textwidth]{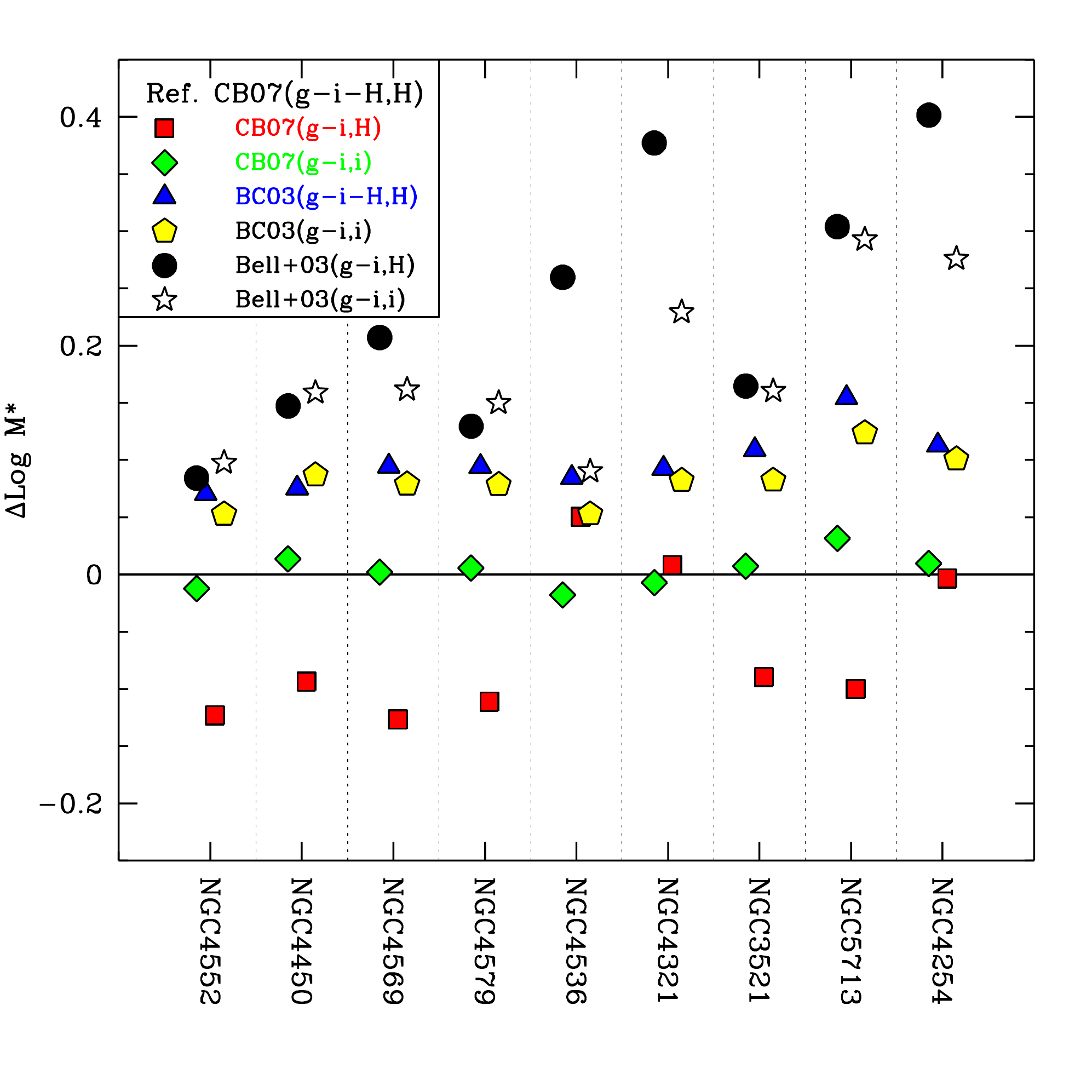}
\caption{Comparison between total stellar masses from resolved maps
  derived using different methods and/or models. For each galaxy the
  logarithmic difference with respect to our fiducial method is
  plotted. Different symbols correspond to different methods/models,
  as indicated in the legend.}\label{fig_res_mass_comp}
\end{figure}
We further explore the relative bias of different stellar mass
estimation methods in Figure \ref{fig_res_mass_comp}. We consider the
total stellar mass of galaxies as given by the integral of the
resolved maps and plot the logarithmic difference with respect to our
reference method (a). In addition to the four methods discussed above
we show the relative bias for other three methods: $\Upsilon_i(g-i)$
based on BC03 models, $\Upsilon_H(g-i)$ and $\Upsilon_i(g-i)$ based on
\cite{bell+03} fitting formulae. Galaxies are sorted in morphological
type, from E to the left to Sc to the right, and different symbols
refer to different methods, according to the figure legend.  We
observe that total mass estimates from $\Upsilon_i(g-i)$ (CB07) is in
excellent agreement with our fiducial method $\Upsilon_H(g-i,i-H)$
(CB07): absolute deviations are $< 0.03$ dex for all galaxies ($0.012$
dex on average) and no bias is seen ($<\Delta \log M_*>=-0.004$ dex).
As already noted on the resolved maps of Figures
\ref{fig_N4321_mass_comp} and \ref{fig_N5713_mass_comp}, using the
optical $g-i$ colour index alone to constrain $M/L$ in NIR bands
provides instead a very poor match to the results of the 2-colour
method: mass estimates are biased low by approximately 0.1 dex for
most galaxies and a substantial scatter is seen.

Mass estimates based on BC03 models give systematically higher values
with respect to CB07: this is not surprising given the lower $M/L$ in
CB07 due to the new TP-AGB prescription and is consistent with Figure
8 of \cite{cimatti+08}. However, by comparing stellar mass estimates
based on BC03 model predictions we note that those based on $H$-band
are systematically higher than those based on $i$-band, as opposed to
the absence of bias observed with CB07 models. This might be an
indication that the role of TP-AGB stars in BC03 models is
under-estimated: in fact, the emission of these stars is relatively
more intense in $H$ than in $i$ band; hence, by under-estimating their
contribution one would conversely over-estimate $M/L$ by a larger
amount in $H$ than in $i$ band.

Mass estimates based on the fitting formulae of \cite{bell+03} are
substantially higher than all mass estimates based on our models, as
expected from Fig. \ref{fig_SPS_c} and \ref{fig_SPS_d}. In particular,
using these fitting formulae we grossly over-estimates the stellar
masses of blue star-forming galaxies, up to by a factor 2.5. We stress
that the difference cannot be due to the new TP-AGB prescription
alone: the \cite{bell+03} formulae rely on BC03 SPS models, yet they
over-estimates stellar masses even relative to our BC03-based methods.
As discussed in section \ref{SPS_sec}, the main reason why our
estimates differ from Bell et al.'s is the different prior
distribution of star formation histories, particularly concerning the
age and the relative importance of bursts. Despite the smaller
dynamical range of $M/L$ in Bell et al.'s models, mass estimates
derived from their prescription based on $i$ and $H$ band respectively
are in reasonable agreement only for 6/9 galaxies; in the other three
cases they disagree by approximately 0.15 dex (40 per cent). This
indicates that Bell et al.'s fitting formulae have, in general, a poor
internal consistency if applied pixel by pixel.

\subsection{Mass estimates from colour maps vs. global
  colours}\label{mass_bias_sec}
\begin{figure*}
\includegraphics[width=\textwidth]{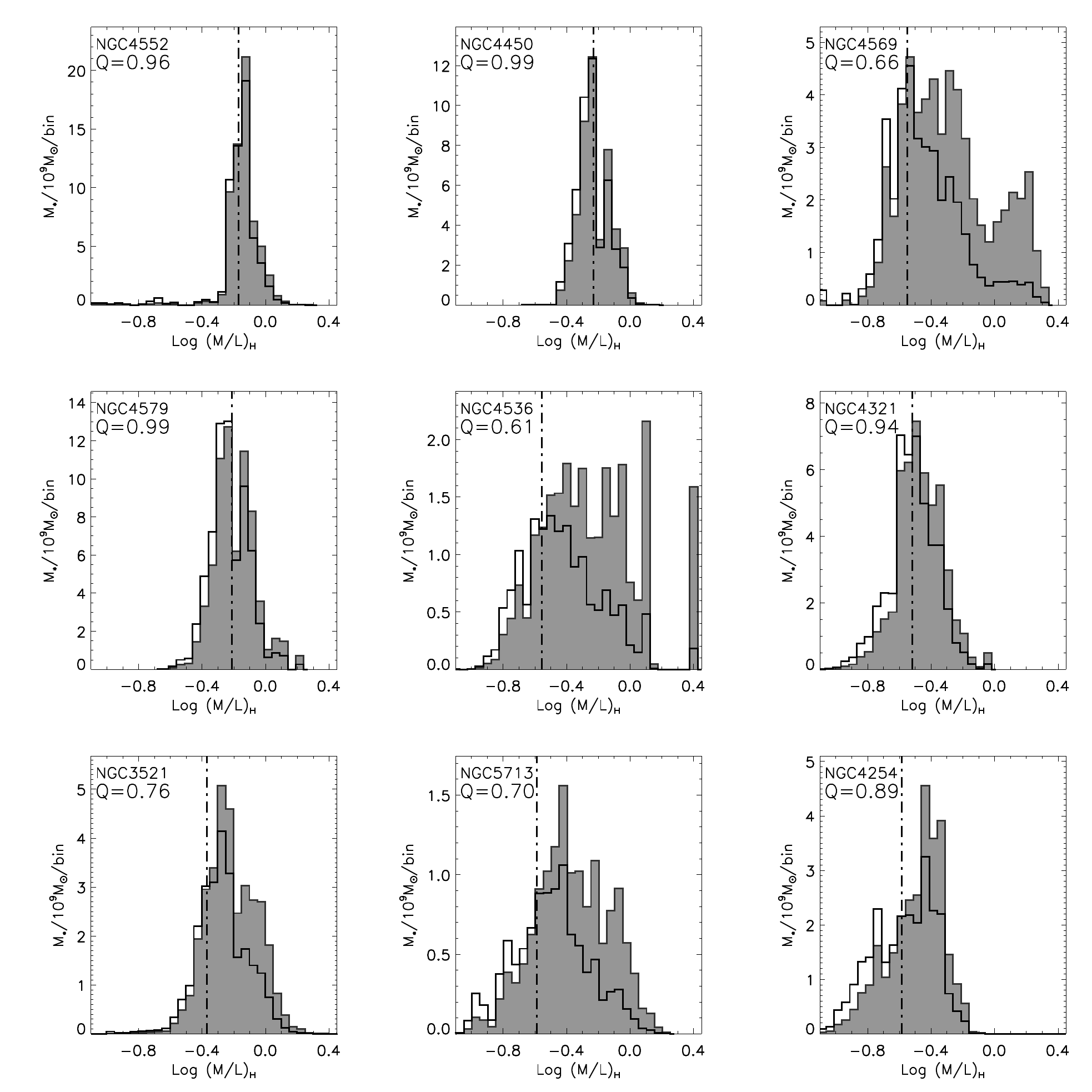}
\caption{Distributions of stellar mass computed using the proper local
  $\Upsilon_H$ (grey histogram) and the global $\Upsilon_H$ as derived
  from global colours (empty histogram), as a function of {\it local}
  $\Upsilon_H$. The dot-dashed line marks the global $\Upsilon_H$. The
  ratio $Q$ between total stellar mass estimated drawing on unresolved
  and on resolved photometry is also reported in each
  panel.}\label{fig_mass_MLdistrib}
\end{figure*}

In this section we investigate the difference in determining the total
stellar mass of a galaxy by integrating resolved stellar mass maps,
like those presented in the previous sections, and by using global
fluxes and colours to estimate $M/L$, as is usually done. More
specifically, we are going to compare
\begin{equation}
  M_{*,\mathrm{resolved}}=\sum_j{f_{H,j} \Upsilon_H[(g-i)_j,(i-H)_j]}
\end{equation}
against
\begin{equation}
  M_{*,\mathrm{unresolved}}=\Upsilon_H[(g-i)_{\mathrm{global}},(i-H)_{\mathrm{global}}]\sum_j{f_{H,j}}
\end{equation}
where $f_H$ is the $H$-band surface brightness and the index $j$
denotes quantities for individual pixels. We call $Q$ the ratio
$\frac{M_{*,\mathrm{unresolved}}}{M_{*,\mathrm{resolved}}}$ and report
this number for each galaxy in the nine panels of
Fig. \ref{fig_mass_MLdistrib}. For 4 of 9 galaxies the ratio $Q$ is
close to 1 within a few per cent, but for the others $Q<0.9$ down to
0.6. This indicates that the same mass estimator drawing on global
colours misses 40 per cent of the mass measured in a resolved map. The
source of this difference is illustrated by the histograms of
Fig. \ref{fig_mass_MLdistrib} and ultimately can be traced to the
strong non-linearity of the relation between colour(s) and $M/L$. For
each galaxy all pixels are binned according to their estimated {\it
  local} $\Upsilon_H$. The gray-shaded histogram represents the mass
in each bin as computed from the local flux and $\Upsilon_H$. The
empty histogram shows the mass contributed by each bin if $\Upsilon_H$
from total fluxes (marked by the vertical dot-dashed line) were
adopted.  As obvious, the grey histogram is above the empty histogram
for all bins where $\Upsilon_H$ is larger than the global value and,
conversely, it is below for bins where $\Upsilon_H$ is smaller.  The
area below the grey histogram gives the stellar mass estimate from
resolved mass maps, while the area below the empty histogram gives the
mass based on global colours. For narrow $\Upsilon_H$ distributions
the difference between the pixels where the global $M/L$ overestimates
the mass and those where it underestimates it almost balances: this is
the case for NGC4552, NGC4450, NGC4579, NGC4321, that all have
$Q>0.9$. For broader distributions in $\Upsilon_H$ we observe two
facts: {\it i)} the global $M/L$ is lower than the mean of the
distribution, because low $\Upsilon_H$ regions dominate the flux and
hence the colours of the galaxy; {\it ii)} the extended high
$\Upsilon_H$ wing is exponentially amplified in the grey histogram
with respect to the empty one. As a result, galaxies with a broad
$M/L$ distribution have large differences between resolved and
unresolved mass estimates, up to approximately 40 per cent. As we can
see comparing Fig. \ref{fig_mass_MLdistrib} and \ref{fig_sample_RGB},
this is especially the case for galaxies with substantive
dust-obscured regions (i.e. NGC4569, NGC4536, NGC3521 and
NGC5713). Dust obscured regions, in fact, contribute a minor fraction
of the flux and influence the global colours only marginally, although
they may conceal a significant amount of stellar mass. In our most
extreme case, NGC\,4536, regions with $\Upsilon_H>1$ contribute only
roughly 7 per cent of the total $H$-band luminosity, but around 20 per
cent of the stellar mass. Such a low impact in terms of luminosity
also implies that these dust obscured regions can affect the global
colours at a level of $\approx 0.1$~mag at most, such that it is {\it
  observationally} impossible to correctly weight them using
unresolved photometry.

\begin{figure}
\includegraphics[width=0.5\textwidth]{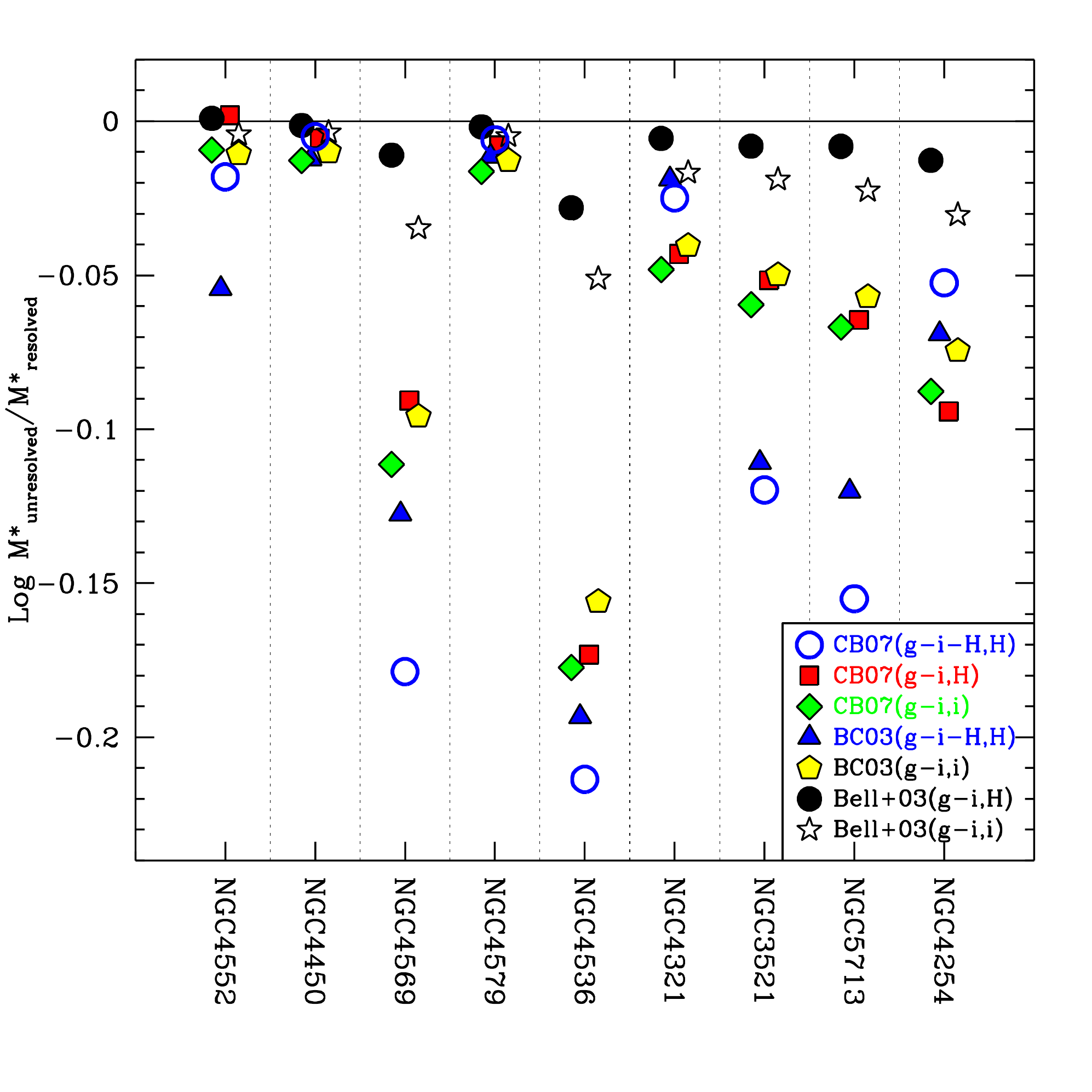}
\caption{Logarithmic differences between total stellar mass estimates
  obtained from unresolved photometry and by integrating resolved
  maps, both based on the {\em same} method/models. Different
  methods/models are shown with different symbols, as shown in the
  legend.  }\label{fig_resunres_mass_comp}
\end{figure}

In Fig. \ref{fig_resunres_mass_comp} we analyze the difference between
unresolved and resolved total stellar mass estimates for different
methods, as indicated in the legend. We plot the logarithmic
difference between unresolved and resolved total stellar mass
estimates (i.e. $\log Q$) both obtained with the same method.  The
dustiest and most irregular galaxies (namely, NGC\,4569, NGC\,4536,
NGC\,3521, NGC\,5713) display the largest differences. They appear
enhanced in two-colour based methods, most likely because these
methods can better disentangle between dust and other stellar
population parameters. For the other five more regular galaxies a
clear trend is observable for all methods: the differences between
resolved and unresolved stellar mass estimates increase going from
early to late types. This is just a consequence of mass differences
being larger for larger pixel colour spread and of colour spread being
larger in later type galaxies (i.e. of early type galaxies being more
uniform).

The dynamical range of models in $M/L$ determines the relative
amplitude of the unresolved estimate bias. For a given model library
and method (one or two colours), this is generally smaller in the NIR
than in $i$ band. The bias is also smaller using Bell et al.'s fitting
formulae, which have a much smaller $M/L$ dynamical range with respect
to our model libraries.

In Section \ref{SPS_bell_sec} we argued that our model libraries are
better suited to describe local scales in terms of SFH and dust
attenuation with respect to the \cite{bell+03} fitting
formulae. However we left the question open whether \cite{bell+03}
might provide better fits to the SED of galaxies as a whole, as they
are less biased toward star formation bursts.  If this is the case, we
must expect Bell et al.'s stellar mass estimates based on unresolved
photometry to agree with our own estimates from resolved maps better
than our estimates based on unresolved photometry do. In order to make
this comparison fair in terms of SPS models, we confront the estimates
based on Bell et al.'s formulae with our BC03-based ones. We find that
mass estimates based on Bell et al.'s $\Upsilon_H(g-i)$ and unresolved
photometry are on average 0.12 dex larger than estimates done by
integrating mass maps based on BC03 $\Upsilon_H(i-H,g-i)$ ({\it
  r.m.s.} 0.1 dex), whereas masses derived with BC03
$\Upsilon_H(i-H,g-i)$ drawing on unresolved photometry are smaller by
0.08 dex on average ({\it r.m.s.} 0.06 dex). This shows that even with
unresolved photometry (and assuming the same SPS models)
\cite{bell+03}'s fitting formulae do not perform better than our
look-up tables do.

\section{Summary and concluding remarks}\label{sec_summary}

In this paper we have developed a method that is capable of
reconstructing resolved stellar mass maps of galaxies from multi-band
optical/NIR imaging, with typical statistical uncertainties of
$0.1-0.15$~dex on local scales. We have realized a Monte Carlo
spectral library of synthetic stellar populations based on the 2007
version of \cite{BC03} code (CB07), which includes a new prescription
to treat the TP-AGB stellar evolutionary phase according to the latest
isochrones by \cite{marigo_girardi07} and
\cite{marigo_girardi08}. Prescriptions to treat dust \`a la
\cite{charlot_fall00} are also incorporated. By marginalizing over all
other parameters we obtain look-up tables that contain median
estimates of $M/L$ in different bands as a function of one or two
optical/NIR colours.

From practical and theoretical considerations we arrive at $g$, $i$
and $H[K_s]$ as a good set of band-passes and express
$\Sigma_{M*}=\Sigma_H \Upsilon_H(g-i,i-H)$: this combination allows to
carefully take young stellar populations and dust obscuration into
account, while avoiding strong H$\alpha$ contamination in HII
regions. We demonstrated that the use of a second colour is required
to determine $\Upsilon_H$ with uncertainties as low as
$0.1-0.15$~dex. Combining $g$ and $i$ bands alone,
$\Sigma_{M*}=\Sigma_i \Upsilon_i(g-i)$, provides a good approximation
to our best method based on $g,~i,~H$ for ``normal'', close to face-on
galaxies. However it may give highly biased results in presence of
very young stellar populations or severe dust extinction, where the
$i$ band flux (7000 $\mathrm{\AA}$) is much more sub-dominant, in the
first case, or more attenuated than in NIR ($1-2~\mu$m), in the second
case.

On the other hand, the flux in the NIR bands appears more sensitive to
the still debated role of TP-AGB stars: old models with shorter-lived
TP-AGB stars over-estimate $M/L$ ratios in $H$-band by $\approx 0.1$
dex (even up to 0.4 dex for young, unextincted stellar populations)
with respect to the current models.

It must be stressed that we account for dust only through its
$4\pi$-averaged extinction \citep[see][]{charlot_fall00}. Although
this assumption is generally reasonable, there are cases where it
fails, such as NGC\,3521 (Fig. \ref{fig_sample_RGB}): it is immediate
to see that the far part of the dusty disk (on the East side) reflects
back the light of the bright inner regions of the small bulge and the
disk. This results in the artificial asymmetry of the reconstructed
mass map of Fig. \ref{fig_sample_mass}. This kind of artifacts are
unavoidable unless a very careful 3D radiative transfer modeling is
performed, which is well beyond the scope of the present mass
reconstruction method.

We have applied our modeling to a small pilot sample of nearby
galaxies, creating pixel-by-pixel maps of the stellar surface mass
density. In general, these maps look quite smooth
(Fig. \ref{fig_sample_mass}), and hence dynamically plausible, with
the prominent spiral arms of young stars and dust greatly reduced.
Detailed comparisons with estimates of dynamical disc massess via
vertical kinematics will enable us to accurately quantify systematics
in our method. We note that preliminary results\footnote{M. Bershady's
  communication in ``Unveiling the mass'' workshop, Queen's
  University, Kingston, Canada, 15-19 June, 2009.} from the DiskMass
survey \citep{verheijen+07} appear in good agreement with the $M/L$
ratios inferred in this work, which are significantly lower than those
derived using the prescriptions of \cite{bell+03}.

Our analysis highlights an important bias in total stellar mass
estimates when spatially unresolved photometric measurements and
colours are used: the stellar mass contribution of dust obscured
regions to the total is severely underestimated from unresolved
photometry, as those regions contribute very little flux and
negligibly affect colours. Mass estimates based on global fluxes can
be biased low by up to 40 per cent. The present sample is too small to
draw conclusions on the consequences that this effect may have for
galaxy stellar mass functions, but we can envisage that resolved mass
estimates may steepen the faint-end slope of the stellar mass
functions as estimated so far. We will address this issue with a
larger and more representative sample in a forthcoming paper of this
series.

After having put the foundations to obtain stellar mass maps in this
paper, in the following papers of this series we will address
questions like: how do structural parameters change going from light
to stellar mass? How can we quantify the bias in unresolved stellar
mass estimates as a function of other observational parameters? How do
SED properties depend on local stellar mass surface density? Can the
inclusion of constraints from longer wavelength improve our mass
reconstruction method?

\section*{Acknowledgments}
We thank Peppo Gavazzi for giving seminal motivation to this work. We
are also thankful to Simon White, Eric Bell, Anna Gallazzi, Matt
Bershady, Kelly Foyle, Elisabete Da Cunha and Paolo Salucci for useful
discussions and comments.

This research has made use of the GOLD Mine Database
(http://goldmine.mib.infn.it).

This work was funded in part by the Marie Curie Initial Training
Network ELIXIR of the European Commission under contract
PITN-GA-2008-214227.

Funding for the SDSS and SDSS-II has been provided by the Alfred
P. Sloan Foundation, the Participating Institutions, the National
Science Foundation, the U.S. Department of Energy, the National
Aeronautics and Space Administration, the Japanese Monbukagakusho, the
Max Planck Society, and the Higher Education Funding Council for
England. The SDSS Web Site is http://www.sdss.org/.\\
The SDSS is managed by the Astrophysical Research Consortium for the
Participating Institutions. The Participating Institutions are the
American Museum of Natural History, Astrophysical Institute Potsdam,
University of Basel, University of Cambridge, Case Western Reserve
University, University of Chicago, Drexel University, Fermilab, the
Institute for Advanced Study, the Japan Participation Group, Johns
Hopkins University, the Joint Institute for Nuclear Astrophysics, the
Kavli Institute for Particle Astrophysics and Cosmology, the Korean
Scientist Group, the Chinese Academy of Sciences (LAMOST), Los Alamos
National Laboratory, the Max-Planck-Institute for Astronomy (MPIA),
the Max-Planck-Institute for Astrophysics (MPA), New Mexico State
University, Ohio State University, University of Pittsburgh,
University of Portsmouth, Princeton University, the United States
Naval Observatory, and the University of Washington.

This research has made use of the NASA/IPAC Extragalactic Database
(NED) which is operated by the Jet Propulsion Laboratory, California
Institute of Technology, under contract with the National Aeronautics
and Space Administration.

\bibliography{mass_maps}
%
%
\onecolumn
\appendix
\section{Model physical parameters in the optical-NIR two-colour
  space}\label{physprop_append}

\begin{figure}
\includegraphics[width=\textwidth]{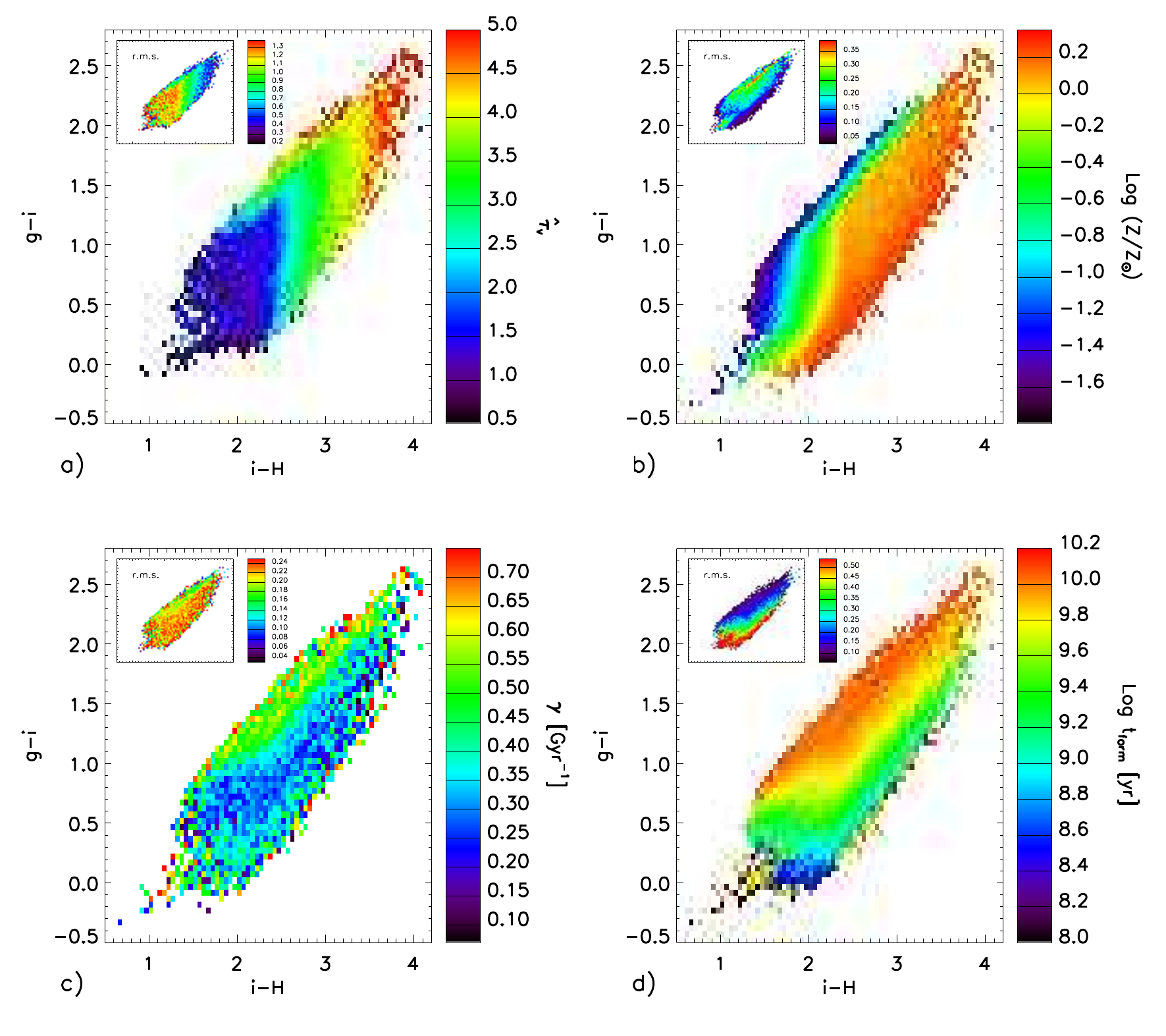}
\caption{The median value of the input physical parameters of our
  library models as a function of $(g-i)$-$(i-H)$. The insets show the
  r.m.s. of the parameter values in each colour-colour bin. {\it Panel
    a)}: the total effective optical depth of the dust in $V$-band,
    $\hat{\tau}_V$; {\it b)}: the stellar metallicity, $Z$; {\it c)}:
    the inverse exponential time scale of the continuous component of
    the SFH, $\gamma$; {\it d)}: the time elapsed since the beginning
    of the SFH, $t_{\mathrm{form}}$.}\label{fig_physprop}
\end{figure}

In this appendix we illustrate how distributed the models of our Monte
Carlo SPS library are in the $(g-i)$-$(i-H)$ colour space in terms of
input physical properties. We consider: the total optical depth of
dust in $V$-band, $\hat{\tau}_V$; the metallicity, $Z$; the inverse
exponential time-scale of star formation, $\gamma$; the time of
formation of the first stars, $t_{\mathrm{form}}$ (taken as look-back
time). Similarly to Fig. \ref{fig_SPS_a}, each of the four main panels
of Figure \ref{fig_physprop} show the median value of an input
parameter for all models that end up in a given colour-colour bin. The
insets show the corresponding r.m.s. We can easily distinguish two
main regimes: {\it blue} models with $g-i\lesssim 1$ and $i-H\lesssim
2.3$, and all the rest. {\it Blue} models have the lowest
$\hat{\tau}_V$ and $\gamma \lesssim 0.4 \mathrm{Gyr}^{-1}$
corresponding to star-formation time scales longer than 2.5 Gyr,
without any obvious dependence on colours and with a uniform large
scatter. As opposed, the models in the {\it blue} regime span the
whole range of $t_{\mathrm{form}}$ and metallicity, with an orthogonal
systematic dependence on colours: older $t_{\mathrm{form}}$ correspond
to redder $g-i$ irrespective of $i-H$, while higher metallicities
correspond to redder $i-H$, almost independent on $g-i$. In the rest
of the colour-colour space the change in colours appears to be largely
driven by the dust, with the other three parameters being decisive
only to determine the most extreme colours (i.e. at the edges of the
distribution).

\section{Powerlaw fits to $M/L$ as a function of one
  colour}\label{plaw_fits_append}
In Table \ref{plaw_fits_tab} we report the parameters of powerlaw fits
to the $M/L$ in different bandpasses as a function of one optical
colour: 
\begin{equation}
\log\Upsilon_\lambda({\rm colour})= a_\lambda + (b_\lambda
\times {\rm colour})
\end{equation}\label{equation:plaw}
This table is meant to provide a direct comparison with table 7 of
\cite{bell+03}.  Power-law fits to the $M/L$ ratios of our CB07-based
models are estimated by the following robust method: in first place
models are binned in colour in intervals of 0.05 mag; for each bin the
median $M/L$ is computed and finally the power-law fit is computed via
weighted linear least squares, where the number of models in each bin
is adopted as weight. Magnitudes in the SDSS bands are meant to be in
the AB systems, while for Johnson-Cousins filters they are expressed
in Vega units.

\begin{table*}
\begin{minipage}{\textwidth}
  \caption{Powerlaw fitting parameters for $\log\Upsilon_\lambda({\rm
      colour})= a_\lambda + (b_\lambda \times {\rm
      colour})$}\label{plaw_fits_tab}
\begin{tabular}{lcccccccccccccc}
  \hline
  \hline
  Colour & $a_g$ & $b_g$ & $a_r$ & $b_r$ & $a_i$ & $b_i$ & $a_z$ & $b_z$ & $a_J$ & $b_J$ & $a_H$ & $b_H$ & $a_K$ & $b_K$ \\
  \hline
  $u-g$&-1.628&  1.360& -1.319&  1.093& -1.277&  0.980& -1.315&  0.913& -1.350&  0.804& -1.467&  0.750& -1.578&  0.739\\
  $u-r$&-1.427&  0.835& -1.157&  0.672& -1.130&  0.602& -1.181&  0.561& -1.235&  0.495& -1.361&  0.463& -1.471&  0.455\\
  $u-i$&-1.468&  0.716& -1.193&  0.577& -1.160&  0.517& -1.206&  0.481& -1.256&  0.422& -1.374&  0.393& -1.477&  0.384\\
  $u-z$&-1.559&  0.658& -1.268&  0.531& -1.225&  0.474& -1.260&  0.439& -1.297&  0.383& -1.407&  0.355& -1.501&  0.344\\
  $g-r$&-1.030&  2.053& -0.840&  1.654& -0.845&  1.481& -0.914&  1.382& -1.007&  1.225& -1.147&  1.144& -1.257&  1.119\\
  $g-i$&-1.197&  1.431& -0.977&  1.157& -0.963&  1.032& -1.019&  0.955& -1.098&  0.844& -1.222&  0.780& -1.321&  0.754\\
  $g-z$&-1.370&  1.190& -1.122&  0.965& -1.089&  0.858& -1.129&  0.791& -1.183&  0.689& -1.291&  0.632& -1.379&  0.604\\
  $r-i$&-1.405&  4.280& -1.155&  3.482& -1.114&  3.087& -1.145&  2.828& -1.199&  2.467& -1.296&  2.234& -1.371&  2.109\\
  $r-z$&-1.576&  2.490& -1.298&  2.032& -1.238&  1.797& -1.250&  1.635& -1.271&  1.398& -1.347&  1.247& -1.405&  1.157\\
  \hline
\\
  \hline
  \hline
  Colour & $a_B$ & $b_B$ & $a_V$ & $b_V$ & $a_R$ & $b_R$ & $a_I$ & $b_I$ & $a_J$ & $b_J$ & $a_H$ & $b_H$ & $a_K$ & $b_K$ \\
  \hline
$B-V$&-1.330&  2.237& -1.075&  1.837& -0.989&  1.620& -1.003&  1.475& -1.135&  1.267& -1.274&  1.190& -1.390&  1.176\\ 
$B-R$&-1.614&  1.466& -1.314&  1.208& -1.200&  1.066& -1.192&  0.967& -1.289&  0.822& -1.410&  0.768& -1.513&  0.750 \\
\hline
\end{tabular}
\end{minipage}
\end{table*}
As already discussed in the text, the slopes of our relations are
significantly steeper than those computed by \cite{bell+03} mainly
because of the different assumptions about the star formation history,
in terms of ages and bursts. For the reddest bandpasses the
differences are even larger due to the new prescriptions for TP-AGB
stars that are incorporated in our models.
\label{lastpage}
\end{document}